Three-fold mechanism of inhibition of SARS-CoV-2 infection by the interaction of the spike glycoprotein with heparin.

Giulia Paiardi*[1,2], Stefan Richter[1], Pasqua Oreste[3], Chiara Urbinati[2], Marco Rusnati[2], Rebecca C. Wade*[1,4].

[1] Molecular and Cellular Modeling Group, Heidelberg Institute for Theoretical Studies (HITS), 69118 Heidelberg, Germany,
[2] Macromolecular Interaction Analysis Unit, Section of Experimental Oncology and Immunology, Department of Molecular and Translational Medicine, 25123 Brescia, Italy,
[3] Glycores 2000 Srl, Milan, Italy.
[4] Zentrum für Molekulare Biologie (ZMBH), DKFZ-ZMBH Alliance and Interdisciplinary Center for Scientific Computing (IWR), Heidelberg University, 69120 Heidelberg, Germany,

*Corresponding authors: Giulia Paiardi, Rebecca C. Wade

Email: giulia.paiardi@h-its.org, Rebecca.wade@h-its.org

**Abstract**

Heparin has been found to have antiviral activity against SARS-CoV-2. Here, by means of sliding window docking, molecular dynamics simulations and biochemical assays, we investigate the binding mode of heparin to the virus spike glycoprotein and the molecular basis for its antiviral activity. The simulations show that heparin binds at long, mostly positively charged patches on the spike, thereby masking the basic residues of the receptor binding domain and of the S1/S2 site. Experiments corroborated the simulation results by showing that heparin inhibits the cleavage of spike by furin by binding to the basic S1/S2 site. Our results indicate that heparin exerts its antiviral activity by both direct and allosteric mechanisms. Furthermore, the simulations provide insights into how heparan sulfate proteoglycans on the host cell can facilitate viral infection. Our results will aid the rational optimization of heparin derivatives for SARS-CoV-2 antiviral therapy.

## Introduction

In the last year, the COVID-19 pandemic caused by severe acute respiratory syndrome coronavirus 2 (SARS-CoV-2) has adversely affected the lives of all people around the world. SARS-CoV-2 is a lipid-enveloped positive-sense RNA virus belonging to the *Coronaviridae* family (1,2,3). Among the SARS-CoV-2 proteins, the spike S glycoprotein (spike) is highly conserved in the *Coronaviridae* family (76% and 96% amino acid residue sequence similarity with SARS-CoV-1 and BatCoV-RaTG13, respectively (1)). It possesses 22 N-linked glycans that likely contribute to protein stability and immune evasion of the virus (4). The spike mediates the entry of the virus into human cells by binding to the angiotensin-converting enzyme 2 (ACE2) receptor (5). The prefusion spike is exposed on the virion surface as a homotrimer. Each spike subunit is composed of two domains, S1 and S2, connected by the S1/S2 junction, and involved in virus attachment and fusion, respectively (5). The S1/S2 junction is a novel feature of the SARS-CoV-2 spike that consists of a multibasic sequence that is cleaved by the host furin protease (1,3,5,6). This cleavage activates the formation of the post-fusion conformation of spike, which is necessary for SARS-CoV-2 entry into host cells (5,6). Intriguingly, the S1/S2 junction is part of a recently identified superantigen motif on the SARS-CoV-2 spike that has been proposed to interact with T-cell receptors and thereby elicit a hyperinflammatory response (7).

Heparan sulfate proteoglycans (HSPGs) on the surface of human host cells are composed of a core protein that harbours multiple polysulfated glycosaminoglycan chains with a similar structure to heparin (8). Experimental evidence suggests an essential role for HSPGs as co-receptors that, by binding spike, favor SARS-CoV-2 attachment to human cells (9,10,11). Although heparin is currently used to treat COVID-19 patients because of its strong anticoagulant activity (12), it has been shown to exert antiviral activity, primarily in its unfractionated state (up to 15kDa corresponding to ~50 monosaccharides), likely due to its ability to compete with HSPGs for binding to the spike (9,10, 11, 13,14).

The binding of spike to heparin or HSPGs is mediated by clusters of positively charged amino acid residues in the spike (hereafter called heparin binding domains, HBDs), and negatively charged sulfate groups on the polysaccharide chains of heparin or the HSPGs (8,9,10). To date, three putative HBDs have been identified in the spike sequence: in the receptor binding domain (RBD), at the S2' TMPRSS2 cleavage site (15) (both present also in SARS-CoV-1, and at the novel S1/S2 furin cleavage site (10) (Fig. 1A).

While the mechanism underlying HSPG-mediated internalization and the antiviral activity of heparin are well understood for many HSPG-dependent viruses (16), it has only recently begun to be investigated for SARS-CoV-2. Indeed, among the various coronaviruses, the SARS-CoV-2 spike is unique in possessing three different putative HBDs that overlap with motifs that have distinct functions (RBD, TMPRSS2 and S1/S2 sites), indicating that spike-HSPG or spike-heparin interactions may have multifaceted effects on SARS-CoV-2 infection. In this context, we here report the results of molecular modeling and simulation along with biochemical assays aimed at revealing how heparin exerts its antiviral effects on SARS-CoV-2 and how HSPGs can instead facilitate virus infection. In particular, we consider the mechanistic consequences of the sequence features that are specific to the SARS-CoV-2 spike for the effects of heparin and HSPGs on viral infection and inflammation.

## Results

- **Long mostly basic patches on the spike glycoprotein accommodate heparin polysaccharide chains**

We considered the inactive closed and active open prefusion conformations of the homotrimeric SARS-CoV-2 spike head, which consists of three subunits: $S_A$, $S_B$ and $S_C$. In the inactive closed conformation, the ACE2 receptor binding face on the RBD is not accessible in any of the three subunits, which we refer to as down-subunits. In the active open conformation, the ACE2 binding site of the RBD is accessible only in the $S_C$ subunit, which we refer to as the up-subunit. To investigate the binding of heparin to spike, we modelled five systems with the spike: (i) in the closed conformation; (ii) in the closed conformation with a single heparin chain bound; (iii) in the closed conformation with three heparin chains bound; (iv) in the open conformation; and (v) in the open conformation with three heparin chains bound.

The model of the SARS-CoV-2 spike head (Fig. 1B) was based on the SwissModel model (available at https://swissmodel.expasy.org/repository/species/2697049) to which we covalently attached 18 N-glycans per subunit (4). To identify continuous positively charged paths on the protein surface at which the long heparin chains could bind, the electrostatic potential of the spike head protein was computed for both the closed and open conformations. Analysis of the electrostatic potential suggests that a long polyanionic heparin chain can follow similar paths on the surfaces of the two conformations of the spike head that differ only in the interactions with the RBD-HBD (Fig. 1B). We used the incremental docking and sliding window method (17) to model complexes of the spike head with heparin chains of 31 monosaccharides (31mer). The docked heparin chains run along mostly basic patches, passing partially through surface grooves, from the S1/S2 basic motif (R682, R683, R685), via the channel between the N-terminal domain of the same spike subunit and the RBD-HBD of an adjacent spike subunit (R346, R355, K356, R357) to the tip of the spike head.

Due to the structural similarity between heparin and the polysaccharide chains of HSPGs, it is expected that the latter also bind along the basic paths identified on the trimeric spike head. By burying these basic regions, heparin can be expected to hinder the binding of HSPGs, reducing the amount of SARS-CoV-2 that can tether to the cell surface, thus decreasing the binding to the ACE2 receptor and hence infection. This is one of the mechanistic explanations that the models provide for the experimental data demonstrating that heparin can diminish SARS-CoV-2 infection (9,10, 11, 13,14). To investigate further mechanisms by which heparin can exert antiviral activity against SARS-CoV-2, we proceeded to carrying out molecular dynamics (MD) simulations of the modelled systems.

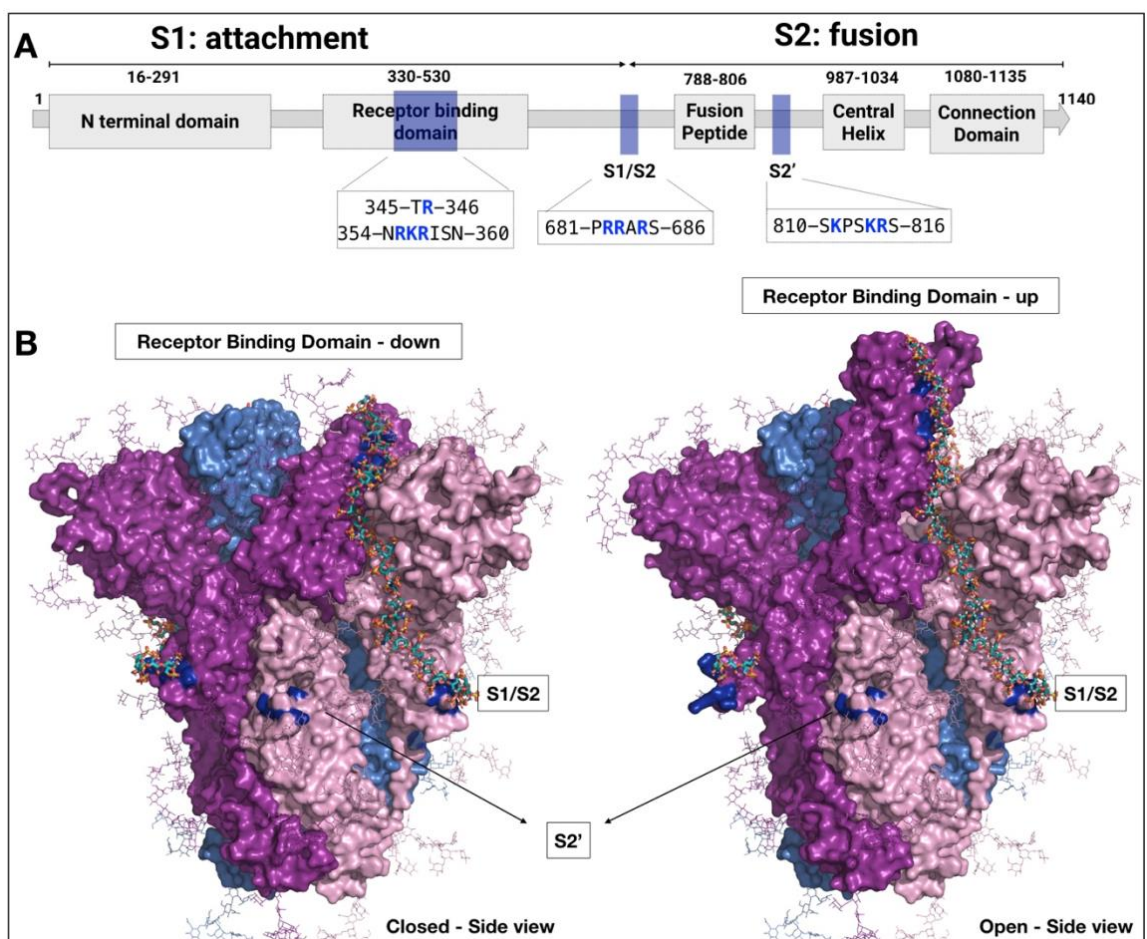

**Figure 1.** *Modeling of spike-heparin interactions.* (**A**) Schematic diagram of the sequence of the SARS-CoV-2 spike (S) glycoprotein head (residues 1-1140), which is composed of S1 and S2 subunits. The boxes along the sequence (Uniprot P0DTC2) show the positions of the main protein domains: N-terminal domain (16-291), receptor binding domain (RBD) (330-530), S1/S2 site (681-686), fusion peptide (788-806), S2' domain (810-816), central helix (987-1034) and connection domain (1080-1135). Three putative heparin binding domains (HBDs) are indicated by blue boxes with their corresponding sequences. The S1/S2-HBD is unique to SARS-CoV-2. (**B**) Side views of the models of the spike head homotrimer before the MD simulations in closed (left) and open (right) conformations with 3 heparin chains bound. Only one complete heparin chain can be seen from this view. The $S_A$, $S_B$ and $S_C$ subunits are shown as surfaces in blue, pink, and magenta, respectively. The dark blue surface patches correspond to the putative HBDs. N-glycans covalently attached to the spike are shown in line representation, colored according to the subunit to which they are attached. The 31-mer heparin chains that span from the S1/S2-HBD to the RBD-HBD are shown in stick representation, colored by element with cyan carbons.

- **Heparin stabilizes the spike glycoprotein in the closed conformation**

The five modelled systems were each simulated in aqueous solution in four replica all-atom MD simulations, each of 1 µs duration (Movies S1-S3). Representative structures obtained after clustering analysis of the spike in closed and open conformations with 3 heparin chains bound show that the heparin chains maintain their alignment along the long positively charged surface patches on the spike during the simulations (Fig. 2A). The spike structures reached convergence within ~2-400 ns in all simulations, as shown by the root mean square deviation (RMSD) relative to each starting structure (Fig. 2B and Fig. S1). Moreover, when in complex with heparin, the spike showed an approximately 1 Å lower

RMSD, indicating that the binding to heparin stabilized the homotrimer structure (Fig. 2B and Fig. S1). Further evidence of the structural stabilization was provided by the root mean square fluctuations (RMSF), which tended to lower values for the RBD, the hinge region associated with the opening of the RBD (see next section for further details), and the S1/S2 site when bound to heparin (Fig. S2, Tab. S1-S3). Accordingly, the RMSD of heparin indicates an induced fit along the trajectory (Fig. 2B and Fig. S1). The complementarity of the interactions and the change in the width of the basic groove in both the closed and open spike structures bound to heparin suggest an induced fit upon binding that results in the well-defined partially grooved basic path where heparin lies (Fig. 2A).

Overall, the H-bond and interaction fingerprint analysis (18) of all the MD simulations show that each heparin chain maintains stable interactions with two adjacent spike subunits in both the closed and open conformations of spike (Fig. 2A, Fig. S3, Tab. S4-S6). Heparin binds through H-bonding interactions (with over 90% occupancy throughout all the MD trajectories) to the basic residues of the S1/S2-HBD in the first subunit and of the RBD-HBD of the second subunit of both the open and closed spike conformations. However, along the heparin path, additional, less specific binding regions, which differ between the open and the closed conformations, were identified in over 75% of the MD frames and these can further stabilize the complexes.

In the closed spike model, the binding of either one or three heparin chains hinders the opening of spike by stabilizing the closed conformation through the simultaneous binding of the RBD of one subunit (residues T345, R346, N354, R355, S359, N360, N450) and the N-terminal domain of the adjacent subunit (residues N165, C166, T167, E169, V171, Q173, F220, N280, N282, T284, T286) (Fig. 2A, Fig.S3, Tab. S4 and Tab. S6). In the closed conformation of spike, the RBDs are essentially strapped into the down-orientation by heparin which prevents spike activation to the open conformation. Finally, heparin binding is also stabilized by hydrogen-bond interactions adjacent to the multibasic S1/S2 site (residues N606, S686, S689, S690) (Fig.S3, Tab.S1 and Tab.S3).

The simulations of spike in the open state do not show any tendency for heparin to induce closure of the spike. However, through the simultaneous binding to the up-RBD of subunit $S_C$ (residues T345, R346, N354, R355, R357, N360) and the N-terminal domain of the adjacent subunit (residues R34, T167, E169, Q173, L176, R190, H207, T208, F220, S221), heparin induces a change in the orientation of the RBD in the up-subunit during the simulations (Fig 2A, Fig. S3, Tab. S5-S6 and next section for further details). As for the closed models, some polar residues near to the multibasic S1/S2 site (N606, Y674, S686, Q689, S690) permanently interact with the heparin chain in the open state.

Notably, the simulations of both open and closed forms of spike with heparin bound suggest an aspecific modulatory effect of N-glycans on the binding between spike and heparin, see Figure S4 and Movies S1-S3.

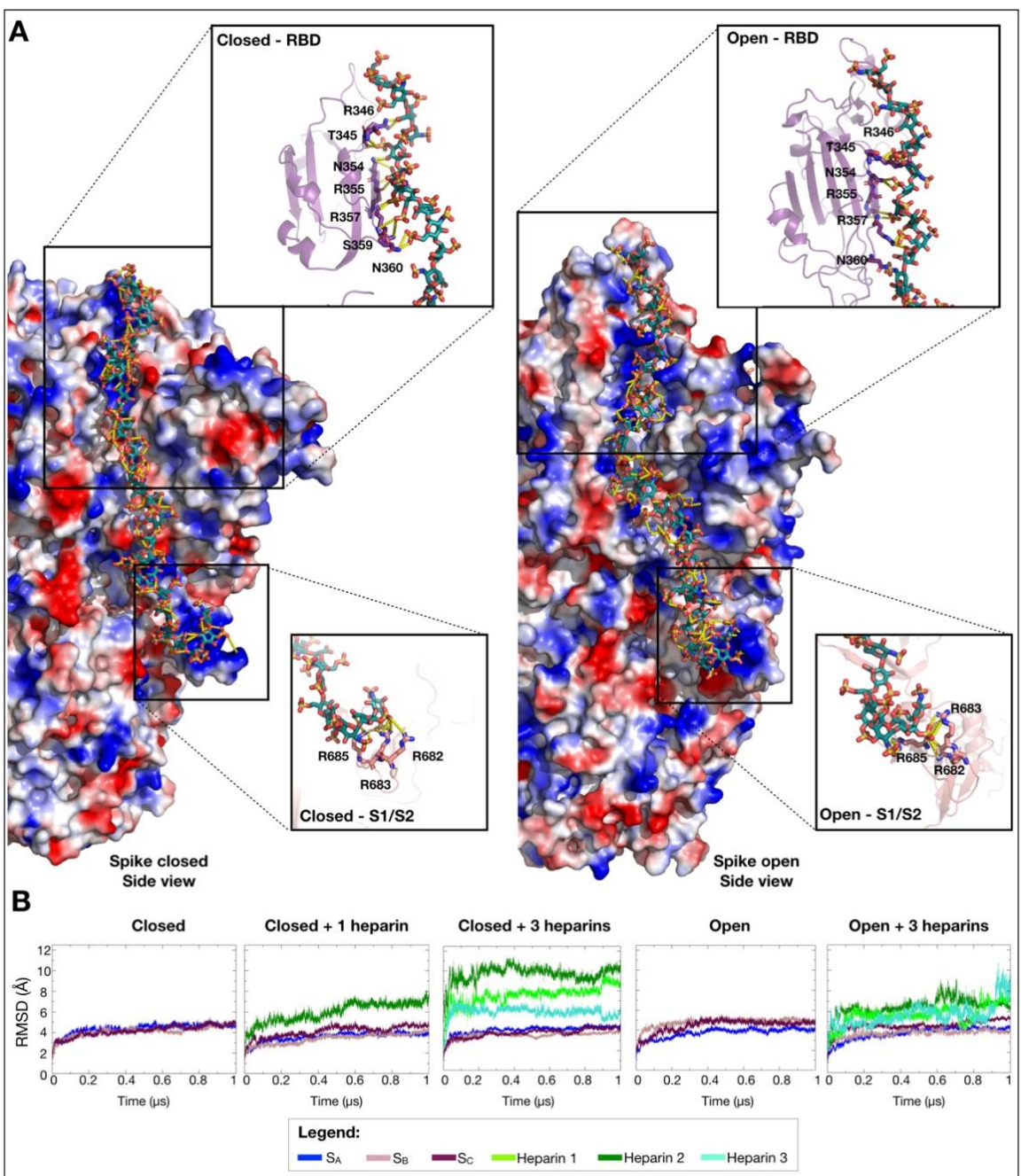

**Figure 2.** *Stability of the spike-heparin complexes. (**A**) Representative closed (left) and open (right) structures obtained after MD simulation of spike bound to three heparin chains are displayed as molecular surfaces with electrostatic potential mapped onto them to show the partially grooved positively charged path occupied by a heparin 31mer. Heparin is shown in stick representation colored by element with cyan carbons. The yellow dashed lines show H-bonds between the spike and heparin. The insets for the closed and open conformations highlight the H-bonding interactions between heparin and the residues in the RBD (T345, R346, N354, R355, N360) and S1/S2 (R682, R683, R685) HBDs shown in stick representation with carbons colored according to the subunits to which they are bound. (**B**) Structural convergence of the spike and heparin chains in the five simulated systems. RMSD versus time is shown for one replica MD simulation for each modelled system. Corresponding plots for all replica MD simulations are shown in Fig.S1.*

- **Heparin masks the spike S1/S2 site**

To assess the ability of heparin to mask the novel furin cleavage site (5,6), which has also been identified as a novel putative HBD (10) and as part of a superantigenic sequence (7) in the spike of SARS-CoV-2, H-bond formation to the site was monitored along the trajectories and surface exposure of the S1/S2 site to the solvent was analyzed for representative structures after clustering (Fig. 3 and Fig. S5, Tab. S4-S7). In the initial modelled complexes, the first monosaccharides of the heparin chain interact with the basic residues of the S1/S2-HBD. Salt-links with R682, R683 and R685 are maintained along all the trajectories (>90% occupancy), indicating strong interactions of the S1/S2-HBD with heparin (Fig. 3A, Tab. S4-S6). The calculated solvent accessible surface area (SASA) of this multibasic site shows the persistence of the interactions during the simulations of the closed spike bound to heparin (Fig. 3B and Fig. S5). Heparin approximately halves the surface exposed in the closed models by binding directly to the S1/S2-HBD and the three heparin chains simultaneously mask all three multibasic sites on the closed spike trimer. In the open conformation, the reduced SASA in the presence of heparin indicates a significant shielding effect, primarily at the S1/S2-HBD of the up-subunit, but with the heparin chains also maintaining a lower shielding level at the S1/S2-HBDs of the down-subunits. This difference could be due to a more favourable arrangement of the basic patches for heparin binding between the up- and down-subunits compared to that between two down-subunits.

To assess the shielding effect of heparin relative to that of spike glycosylation, the SASA of the S1/S2-HBD calculated for the representative clusters was decomposed into the area exposed without consideration of the N-glycans and heparin sugars, the area exposed accounting for the N-glycans, and the area exposed accounting for both the N-glycans and heparin (Fig. 3C and Tab. S7). In agreement with the previous analysis, these calculations indicate that heparin directly binds to the S1/S2-HBDs, halving the exposed surfaces in both the closed and open conformations. Again, for the open conformation, the shielding effect exerted by heparin on the S1/S2-HBDs of the two down-subunits is lower than for the same site of the up-subunit. Moreover, the decomposition shows that the N-glycans of spike make little contribution to the shielding of the multibasic sites.

In summary, when comparing the binding of one or three heparin chains to the spike homotrimer, all the data point to the ability of heparin to occupy and shield the S1/S2 site without a significant shielding contribution of the spike N-glycans. In addition, our simulations demonstrate a key role of the S1/S2 site in the binding of heparin and indicate its involvement in binding to HSPGs. The surface area analysis shows that the S1/S2-HBD is exposed to the solvent and available to interact with heparin or HSPGs, supporting the hypothesis that this HBD, which is specific to SARS-CoV-2, may contribute to the increased affinity of SARS-CoV-2 spike for heparin and HSPGs compared to previous coronavirus strains. Moreover, heparin could exert its antiviral activity by masking this site, preventing cleavage by furin and hindering the activation of the post-fusion conformation of the spike glycoprotein. Interestingly, we observed that, in the closed spike model, a single heparin chain occupies the S1/S2-HBD of only one subunit. This suggests that, to exert its full antiviral activity, heparin needs to be administered at doses high enough to saturate all three S1/S2-HBDs of the spike trimer, otherwise some of them will still be available for tethering to the HSPGs on the host cell surface.

- **Heparin prevents cleavage at the spike S1/S2 site by furin**

To confirm the hypothesis from our simulations, we measured the effect of heparin on the cleavage of a spike fragment containing the S1/S2 site. As shown in Fig. 3D, heparin effectively inhibits spike cleavage by furin. The effect is specific since heparin alone does not interfere with the colorimetric assays. To assess whether the inhibition of furin cleavage is dependent on the binding of heparin to the S1/S2 basic site, heparin was pre-incubated with a substrate-immobilized spike fragment, unbound heparin was removed by PBS wash and then furin was added. In these experimental conditions, heparin prevents spike cleavage by furin, indicating that it physically associates with the immobilized spike fragment. On the other hand, a 2M NaCl wash, which is known to disrupt heparin/protein interactions (19,20), completely detached heparin from the spike fragment, restoring cleavage by furin (Fig. 3D). The capacity of heparin to bind spike and to inhibit SARS-CoV-2 infection is dependent on its length and degree of sulfation (13,14). We therefore compared the inhibitory potency of unfractionated heparin (13.6 kDa, ~ 46 monosaccharide units), a low molecular weight heparin (4.0 kDa, ~ 13 monosaccharide units), and unsulfated K5 polysaccharide (30kDa, ~150 monosaccharides). As shown in Fig. 3E, unfractionated heparin exhibits a higher furin inhibitory potency than the low molecular weight heparin, whereas the unsulfated K5 polysaccharide does not prevent furin cleavage. These results demonstrate that, for binding to the spike fragment and protecting it from furin cleavage, the length of the heparin chain is of importance and the presence of sulfated groups is necessary.

These experiments corroborate the hypothesis that heparin exerts its antiviral activity in a length- and dose-dependent manner by masking the S1/S2 site on spike, thereby inhibiting furin cleavage and the formation of the post-fusion conformation. These results also suggest that, by binding to the S1/S2 site, heparin could affect the superantigenic character of the spike by preventing its binding to TCR cells (7).

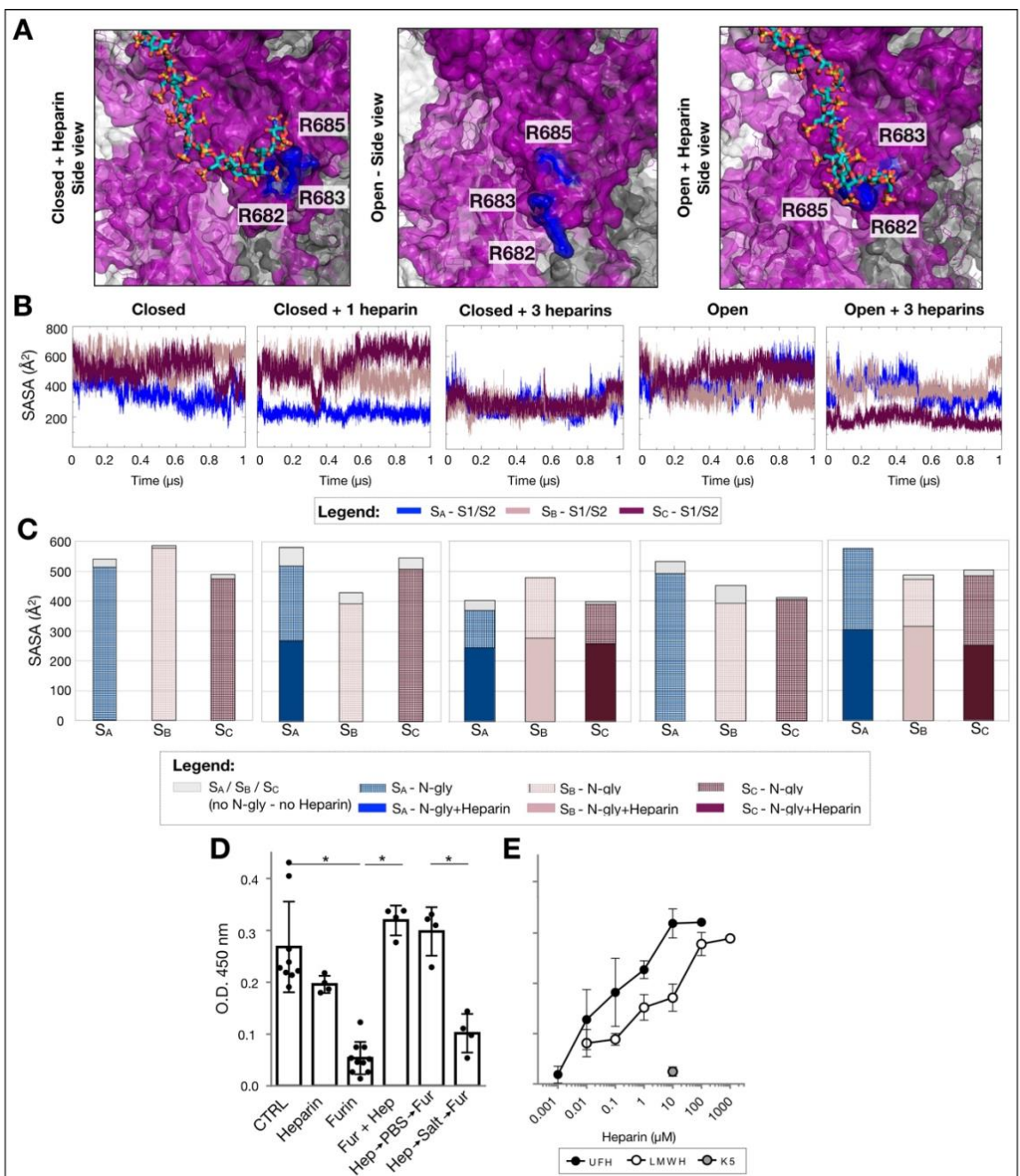

**Figure 3.** *Interaction of heparin with the spike S1/S2-HBD showing shielding of the furin cleavage site by heparin in closed and open spike conformations and inhibition of furin cleavage by heparin. (**A**) The spike S1/S2 site in closed (left) and open (middle, right) representative conformations is shown with and without heparin. The molecular surfaces of the $S_C$ and $S_B$ subunits are shown in magenta and grey, respectively. Basic residues of the S1/S2-HBD are shown in blue and labelled. N-glycans are shown in line representation colored according to the corresponding subunit. Heparin is shown in stick representation colored by element with cyan carbons. (**B**) SASA of the S1/S2-HBD plotted as a function of time for one trajectory for each simulated system: closed, closed with 1 heparin chain, closed with 3 heparin chains, open and open with 3 heparin chains. Corresponding plots for the replica trajectories are shown in Fig. S5. (**C**) Average SASA of the S1/S2-HBD computed for the cluster representatives in the MD simulations for different factors contributing to the burial of the S1/S2 site surface. Standard deviations are reported in Tab. S7. (**D-E**) Effect of heparin on the cleavage of spike by furin. (**D**) Wells coated with a peptide fragment containing*

*the S1/S2 basic site of spike were: left untreated (CTRL), incubated with 13.6 kDa heparin (10 μM) alone or exposed to furin (25 ng/well) in the absence or presence of heparin (10 μM). In addition, the substrate-immobilized spike fragment was pre-incubated with heparin (100 μM) for 10 min., washed three times with PBS or PBS containing 2 M NaCl (salt) and then exposed to furin. Single measurements are reported in plot and in Tab. S8. (**E**) Wells coated with the spike fragment were incubated with furin in the absence or presence of 13.6 KDa unfractionated heparin (UFH), 4.0 kDa low molecular weight heparin (LMWH) or unsulfated K5 polysaccharide (K5) at the indicated concentrations. At the end of the incubations, spike cleavage was measured as described in the Material and Methods section. Each value is the mean ± standard deviation of three to ten repetitions. \*p<0.001, one-way ANOVA. The list of the single measurements for each experiment condition are reported in Tab. S8-S9.*

- **Heparin allosterically affects the hinge region of the RBD and directly interacts with the basic residues of the RBD-HBD**

To assess the effect of heparin binding on the RBD, we analyzed the stability of the hinge region associated with the activation of the RBD, the exposure of the up-RBD (on subunit $S_C$) along the trajectory, and the shielding by heparin of the residues of the RBD involved in the interaction with ACE2, hereafter referred to as the receptor binding motif (RBm) (21).

From a comparison between the crystal structures of the spike in closed and open conformations and the RMSFs in the simulations (data not shown), we identified residues 527-PKK-529 as the hinge region responsible for the conformational change that induces the opening of the spike protein. Importantly, no direct interactions were observed to occur between these residues and heparin (Tab. S4-S6), prompting us to investigate possible allosteric effects induced in this region by the binding of heparin to spike. For this purpose, we calculated the RMSD of the hinge region of each subunit along the trajectory and performed dihedral principal component analysis (dPCA) for this region. As shown in Fig. 4A-B and Fig. S6-7, the closed spike shows structural stability and limited structural variability of the three hinge regions along all the simulations. The closed conformation with one heparin bound shows structural stabilization of the two subunits that are directly involved in the interaction with heparin. However, the $S_C$ subunit, which does not directly interact with heparin, shows a higher RMSD (Fig. 4A and Fig. S6), a higher RMSF (Fig. S2, Tab. S2), and more structural variability (Fig. 4B and Fig. S7), suggesting a compensatory effect due to the movement of the hinge region to maintain the closed conformation. Interestingly, RMSD plots and dPCA of the hinge region of the closed-$S_C$ subunit in the complex with three heparins bound shows a restored stability of the hinge region, indicating the ability of heparin to stabilize the closed conformation of spike and hinder the opening of the RBD. For the open spike, the binding of three heparin chains results in the subunits $S_A$ and $S_B$ with down-RBDs having structurally stable hinge regions whereas the hinge region of the $S_C$ subunit, which has the up-RBD, has an increased RMSD (Fig. 4A and Fig. S6) and altered sampling of the conformational space (Fig. 4B and Fig. S7), indicating a perturbing effect of heparin on the opening of the spike conformation which could affect host-cell receptor binding.

To evaluate if the induced fit promoted by heparin causes the masking of the ACE2 binding residues in the RBm, we calculated the SASA of these residues along the trajectory and their accessibility in the cluster representatives. Both the analyses show that the heparin chains do not significantly shield the residues of the RBm (Fig. 4C-D, Fig. S8 and Tab. S10). On one hand, these data suggest that the heparin chains act indirectly on these domains through an induced fit mechanism as described above. On the other hand, these data are consistent with the ability of spike to simultaneously bind HSPGs and ACE2 on the host cell surface (9,11).

      Finally, to obtain further insights into the effect of heparin on the open RBD, we performed essential dynamics (ED) analysis (Fig. 4E and Fig. S9). The analysis on the closed conformation shows an overall stabilization of the spike by heparin without significant effects on the RBDs (data not shown). ED analysis on the open spike conformation reveals that the binding of heparin results in a different direction of the motion of the RBm loop (residues 472-489) in the up-subunit described by the first eigenvector (Fig. S9). Despite different starting conformations and independent sampling, the up-subunits consistently show this difference across all the replica simulations. These differences suggest that the presence of heparin (or HSPGs) could affect the motion of the RBD, possibly having a gating effect on host-cell receptor binding.

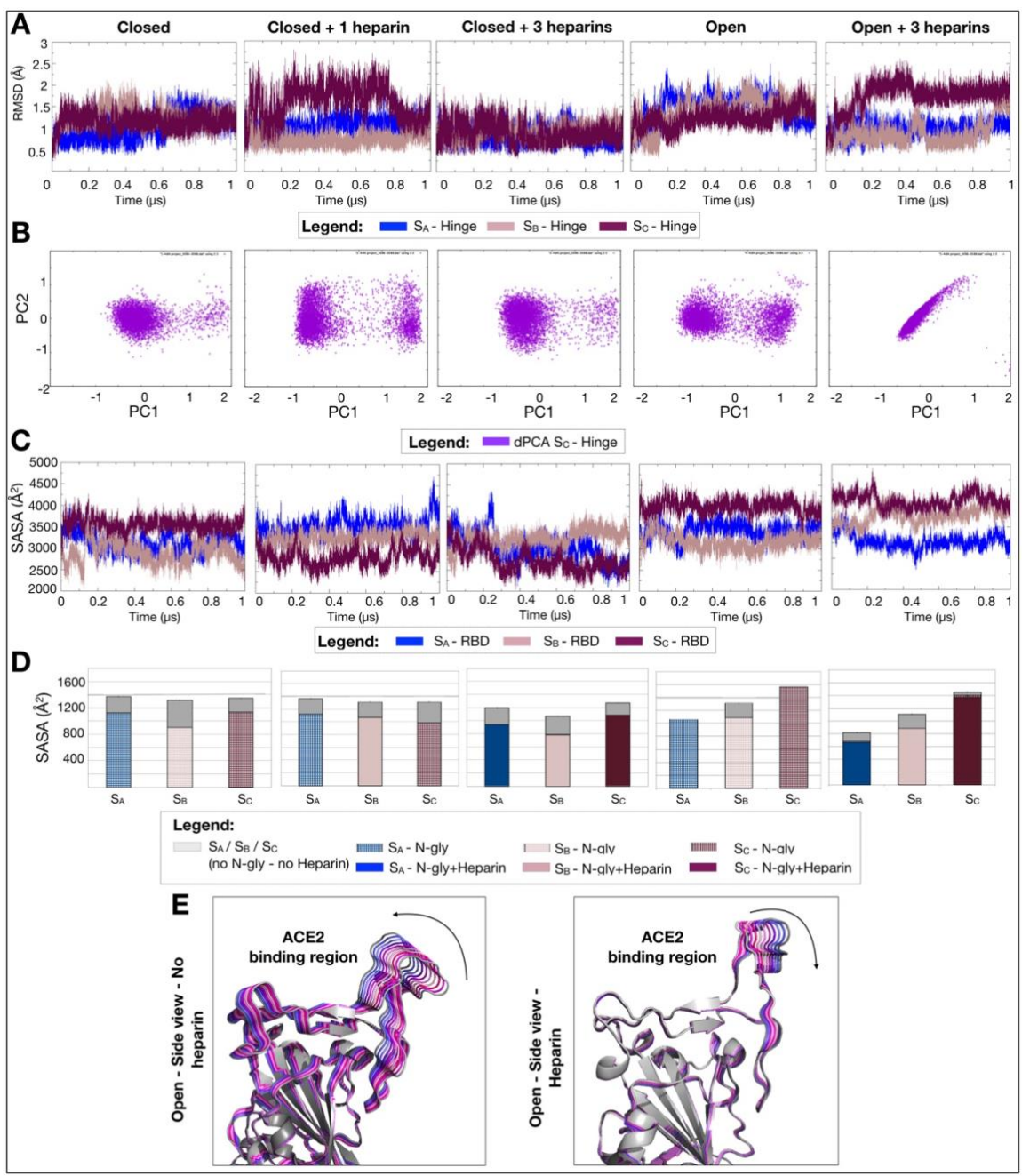

**Figure 4.** *Changes in the conformational flexibility of the spike hinge region (residues 527-529) upon heparin binding and the effect of heparin on the motion of the spike RBD and the accessibility of the receptor binding motif (RBm).* (**A**) *RMSD vs. time and* (**B**) *dPCA of the hinge region for a trajectory for each simulated system showing that heparin binding results in structural stabilization of the hinge region of the closed spike and changes the motion of the hinge region in the open spike. The SASA of the RBm along the corresponding trajectories and in the most representative cluster for each system is reported in* (**C**) *and* (**D**), *respectively, and shows that heparin binding does not reduce the exposure of the RBm (standard deviations are reported in Tab. S10).* (**E**) *The change in motion of the up-RBD induced by the presence of heparin is shown by the superimposition of 10 conformations extracted at equal time intervals along the trajectories (from magenta to blue) and projected onto the first essential dynamics eigenvector without (left) and with (right) heparin. The RBD is shown in cartoon representation and heparin is omitted for ease of*

*visualization. Corresponding plots for the replica trajectories are shown in Figs. S6-S7-S10. See text for details.*

## Discussion

Experiments have shown that HSPGs are indispensable for SARS-CoV-2 infection (9,11), and that heparin binds to the SARS-CoV-2 spike glycoprotein, exerting an antiviral effect (9,10,11,13,14). Importantly, it has been demonstrated that unfractionated heparin has a 150-fold higher antiviral effect against SARS-CoV-2 than low molecular weight heparin (LMWH) (13). Until now, molecular models to investigate the binding of heparin or HSPGs to spike have been limited to models of short heparin chains ($\simeq$6-8 monosaccharides) binding to the basic domain of the spike RBD without N-glycans (9,10,14). However, to understand the antiviral effect of unfractionated heparin and the role of HSPGs, it is necessary to model the binding of long polyanionic chains to the spike head. This is a challenging task due to the length, variable sulfation pattern and flexibility of the polysulfated glycosaminoglycan chains, and because of the large size and flexibility of the spike head for which some regions and N-glycans are structurally poorly defined (4,5). Therefore, considering all available experimental data to build high-quality initial models, we first employed our incremental docking and sliding window method (17) for docking 31mer heparin chains to models of the spike head glycoprotein. We then performed multiple microsecond MD simulations to refine the models of the spike head with zero, one or three heparin chains bound to study the dynamic effects of heparin binding. Despite a total simulation length of over 20 microseconds, the sampling of the configurational space of the spike-heparin systems was inevitably incomplete. Nevertheless, these simulations provided sufficient configurational sampling to allow us to explore the main dynamic features associated with the predicted heparin binding modes and thus, to identify mechanisms by which long heparin chains exert their antiviral activity and HSPGs can act as co-receptors.

We identified three key mechanisms by which heparin exerts its antiviral activity (see Figure 5): (i) heparin directly competes with HSPGs for the same binding sites on the spike head, burying the same basic surface regions and hindering the binding to HSPGs of both the closed and open conformations of the spike head; (ii) heparin masks the S1/S2 multibasic site (unique to SARS-CoV-2), preventing the cleavage by furin and the activation of the prefusion conformation of the spike, as well as modulating the triggering of the hyperinflammatory response due to the binding to T-cell receptors of this superantigenic site (7); (iii) heparin masks the basic residues of the RBD and allosterically acts on the hinge region that is suggested to be key for the opening of the spike and the movement of the RBD to expose the ACE2 receptor binding face. Notably, both the direct and the allosteric mechanisms require long heparin chains. To validate the results of our simulations, we carried out an enzymatic assay that showed that heparin inhibits cleavage of a spike fragment containing the S1/S2 site by furin in a dose-, length- and sulfation-dependent manner and that this binding was effectively due to the binding and masking of the S1/S2 basic site by heparin.

Based on our MD simulations with heparin and on the structural similarity between heparin and the heparan sulfate chains of HSPGs, we can infer that: (i) HSPGs bind along the same paths on the surface of spike occupied by heparin; (ii) HSPGs do not directly induce the activation of the closed spike, suggesting that their role in viral infection may be solely to increase the concentration of the virus on the host-cell surface. However,

HSPGs may also mediate the activation of the closed spike by exposing it to the human ACE2 receptor on the host cell. HSPGs may also favor the formation of a ternary complex (9) with the open spike and the ACE2 receptor by binding the basic residues of the RBD of the up-subunit while not masking the RBm. Notably, both the open and closed spike models with heparin bound suggest a modulatory effect of N-glycans on the binding between spike and heparin or HSPGs. Finally, emerging SARS-CoV-2 spike glycoprotein variants have demonstrated increased infectivity (22). We cannot exclude that this capability could in part be due to mutations near to the basic domains mentioned above that increase binding to HSPGs.

In conclusion, our models and simulations suggest one direct and two allosteric mechanisms for the antiviral activity of heparin against SARS-CoV-2 and provide insights into how HSPGs can facilitate viral infection. They thus provide a basis for the rational optimization of therapeutic heparin derivatives against SARS-CoV-2.

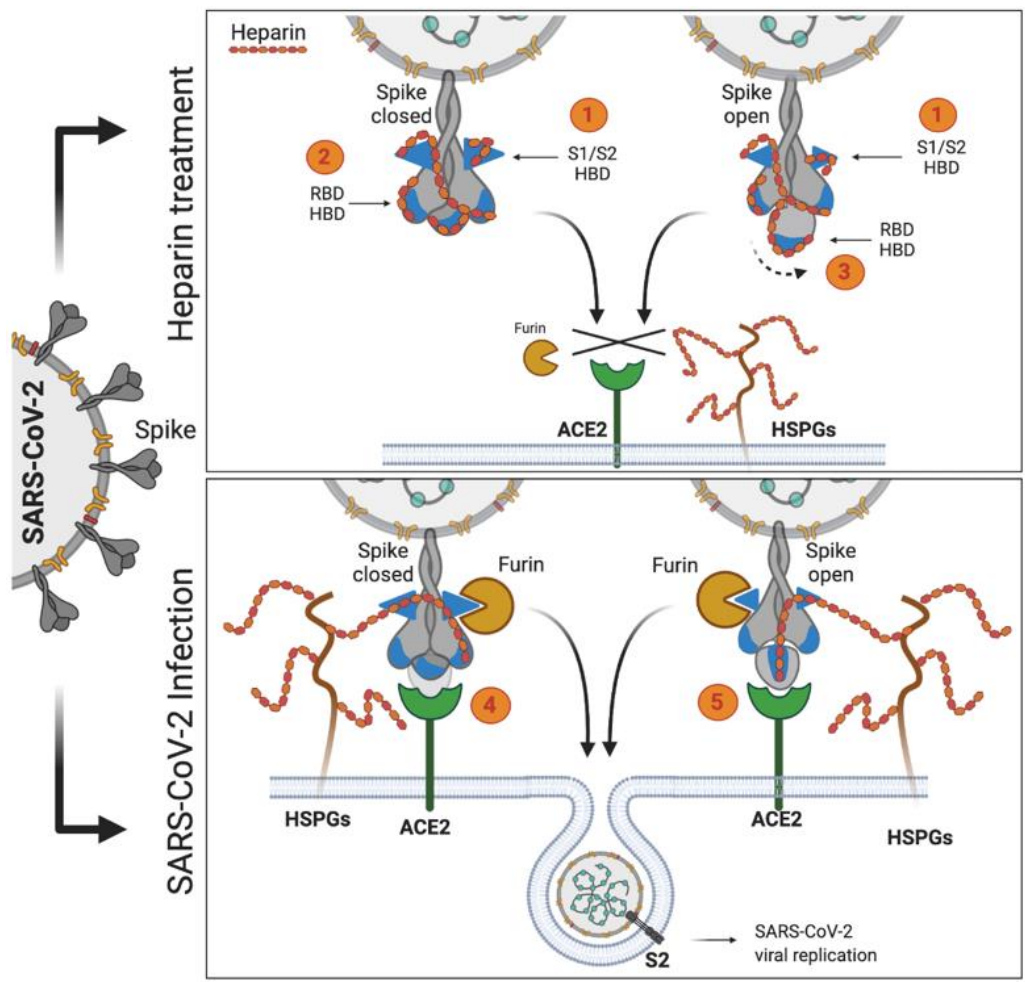

**Figure 5.** *Schematic diagram showing the proposed mechanisms by which heparin and HSPGs can affect SARS-CoV-2 infection of a host cell. The results of the molecular dynamics simulations and biochemical assays suggest the following mechanisms: Heparin exerts antiviral activity by hindering the binding of both closed and open conformations of spike to HSPGs. Heparin binds the S1/S2-HBD of spike in both the conformations, preventing furin cleavage (1), hinders the opening/activation of the closed spike (2), and acts allosterically on the hinge region associated with the movement of the RBD, while directly masking the RBD-*

*HBD (3). Moreover, based on the structural similarity between heparin and HSPGs, we expect that HSPGs are able to bind both the closed (4) and open (5) conformations of spike, and, in the presence of the ACE2 receptor, favor the activation of the closed spike and its interaction with the ACE2 receptor (4), and subsequent furin cleavage and SARS-CoV-2 infection. However, we cannot exclude a possible ACE2-independent SARS-CoV-2 internalization mediated by HSPGs binding to the closed spike conformation.*

## Material and Methods

**Modelling of the systems.** The initial models of the SARS-CoV-2 spike head protein in closed and open conformations were taken from the SwissModel website (https://swissmodel.expasy.org/repository/species/2697049) and were based on two structures determined by cryo-electron microscopy: PDBid 6ACC (seq. identity 76.47%) and PDBid: 6ACD (seq. identity 76.47%), respectively (23). The spike models were completed by adding 18 N-glycans per subunit, covalently attached in accordance with the experimentally determined glycomic profile (4), using the Glycam web (http://glycam.org/) (24). Standard protonation states were used. The APBS electrostatics plugin in Pymol (25) was used to compute the electrostatic potential surface calculated with PDB2PQR (26) at neutral pH using PROPKA (25) and the AMBER ff14SB force (28). 31mer heparin chains spanning from the S1/S2 multibasic site to the RBD-HBD were modelled using the incremental docking and sliding window method developed by Bugatti and co-workers (17) using Autodock 4.2 (29). Five model systems were generated: the closed spike with zero, 1 or 3 heparin chains, and the open spike with zero or 3 heparin chains.

**All-atom Molecular Dynamics (MD) simulation**. The Amber20 package (30) was used to carry out the simulations. Parameters for the spike were assigned with the ff14SB (31) and GLYCAM-06j (32) force fields. Heparin was parameterized following the method published by Bugatti et al. (17). All the glycoprotein models were placed in a periodic cubic water box using the TIP3P water model (33) with 10 Å between the solutes and the edges of the box. $Na^+$ and $Cl^-$ ions were added to neutralize the systems and to immerse them in solvent with an ionic strength of 150mM.

Each system was energy minimized in 14 consecutive minimisation steps, each of 100 steps of steepest descent followed by 900 steps of conjugate gradient, with decreasing positional restraints from 1000 to 0 kcal/mol $A^2$ on all the atoms of the systems excluding waters, counterions and hydrogens, with a cut-off for non-bonded interactions of 8 Å. The systems were then subjected to two consecutive steps of heating, each of 100000 steps, from 10 to 100 K and from 100 to 310 K in an NVT ensemble with a Langevin thermostat. Bonds involving hydrogen atoms were constrained with the SHAKE algorithm (34) and 2 fs time step was used. The systems were then equilibrated at 310 K in 4 consecutive steps of 2.5 ns each in the NPT ensemble with a Langevin thermostat with random velocities assigned at the beginning of each step. For each system simulated, 4 independent replica production runs following the same protocol as for the equilibration, were carried out starting from restart files chosen randomly from the last 5ns of equilibration. During the MD simulations, a cutoff of 8 Å for the evaluation of short-range non-bonded interactions was used and the Particle Mesh Ewald method was employed for the long-range electrostatic interactions. The temperature was kept constant at 310 K with a Langevin thermostat. Coordinates were written at intervals of 100 ps.

Production simulations were carried out on the Marconi100 accelerated cluster (https://www.hpc.cineca.it/hardware/marconi100) which is based on the IBM Power9 AC922 CPU architecture with each accelerator having four Volta V100 NVIDIA GPUs. Each simulation was carried out on a single GPU

**Analysis of MD simulations.** MD trajectories were analysed using CPPTRAJ from AmberTools20 (30) and molecular graphics analysis was performed using Visual Molecular Dynamics (VMD) (35).

- *Cluster analysis* of the structures was carried out using CPPTRAJ (30) for the last 100 ns of all the trajectories considering backbone C-alpha atoms for the protein residues and all the carbon, oxygen, sulfate and nitrogen atoms for the heparin chains. N-glycans were excluded from the analysis. The hierarchical agglomerative (bottom-up) approach was used with a minimum distance between the clusters of 3.0 Å and the distance between clusters defined by the average distance between members of two clusters (Fig. 2A).

- *Hydrogen bond (H-bond) analysis* was performed using CPPTRAJ (30) along all the trajectories for frames at intervals of 10 ns (coordinated collected every 100 ps collected with a stride of 100 frames) and setting 3.5 Å as the upper distance for defining a H-bond between heavy atoms. All the atoms including the hydrogens of the systems were considered (Fig. 2A, Fig.S3, Tab.S1 and S2).

- *Interaction fingerprint analysis (IFP)* was performed using the MD-IFP python scripts (18) (https://github.com/HITS-MCM/MD-IFP). The interactions between the spike and the heparin chains were computed along all the trajectories for frames at intervals of 10 ns (coordinates were written every 100 ps and then analyzed with a stride of 100 frames) (Tab.S3, IFP-model02, IFP-model03, IFP-model05).

- *Root mean square fluctuations (RMSF)* were calculated using CPPTRAJ (30) for all C-alpha atoms of the individual spike subunits - $S_A$, $S_B$, $S_C$ - and for all the carbon, oxygen, sulfate and nitrogen atoms of the heparin chains (Fig.S2).

- *Root mean square deviations (RMSD)* were calculated using CPPTRAJ (30) for all C-alpha atoms of the individual spike subunits - $S_A$, $S_B$, $S_C$ - and for all the carbon, oxygen, sulfate and nitrogen atoms of the heparin chains (Fig. 2B and Fig.S1). The RMSDs of the hinge regions were calculated for the C-alpha atoms of residues 527-529 for $S_A$, $S_B$, and $S_C$ separately (Fig. 4A and Fig.S5).

- *Solvent Accessible surface area (SASA)*. Two separate SASA analyses were carried out for the $S_A$ $S_B$ and $S_C$ subunits separately: along the trajectory using CPPTRAJ (30) and for the most representative clusters using NACCESS (36). In both the analyses, the van der Waals radius of the solvent probe was assigned a value of 1.4 Å. For the analysis of the S1/S2-HBD site, residues 682-685 were considered (Fig.3B-C, Fig.S4, Tab.S4). In the case of the receptor binding residues, all the residues of the RBD (residues 330-530) were considered along the trajectory but only the RBm residues (K417, G446, Y449, Y453, L455, F456, A475, F486, N487, Y489, Q493, G496, Q498, T500, N501, G502, Y505) suggested by Lan and co-workers (21) for the representative clusters (Fig.4C-D, Fig.S7 and Tab.S5).

- *Dihedral Principal Component Analysis (dPCA)* was performed using CPPTRAJ (30). The dihedral covariance matrix and the projection were calculated for the backbone phi/psi angles of residues 527-529 of the $S_C$ monomer. The first four eigenvectors and eigenvalues were extracted and the first two principal components were plotted for all of the systems.

All the systems were transformed into the same principal component space to evaluate the variance across the replicas (Fig. 4B and Fig. S6).

- *Essential Dynamics (ED)* analysis was performed using Principal Component Analysis (PCA) of the unbiased MD simulations. PCA was performed along all the trajectories individually with CPPTRAJ (30). The principal modes of motion were visualized using VMD. The first normalized eigenvectors for model04 and model05 were plotted along the trajectory and the direction of motion was defined by visual inspection (Fig. 4E and Fig.S8).

**Reagents.** Human recombinant furin was obtained from OriGene Technologies Inc. Rockville, MD, USA. Conventional heparin (13.6 kDa) was obtained from a commercial batch preparation of unfractionated sodium heparin from beef mucosa (Laboratori Derivati Organici S.p.A., Milan, Italy) purified to remove contaminants according to established methods (37). The purity was higher than 95% as assessed by electrophoresis in acidic buffer (38), uronic acid quantitative determination (39), and high-performance liquid chromatograph analysis (37). The $^{13}$C NMR spectrum measured following Casu et al. (40) showed 78% N-sulfate glucosamine, 80% 6-O-sulfate glucosamine, and 59% 2-O-sulfate iduronic acid. Low molecular weight (LMW) beef mucosal heparins (4.0 kDa) were obtained by controlled nitrous acid degradation of unfractionated heparin as described in (40,41). The capsular *E. coli* K5 polysaccharide (30,0 kDa) has the same structure [(→4)-β-D-glucuronic acid-(1→4)-α-D-N-acetyl-glucosamine-(1→)]$_n$ as the heparin precursor N-acetyl heparosan and was prepared as described in (42). To assess their integrity, the polysaccharides were re-analyzed before the beginning of the experiments (data not shown).

**Furin cleavage assay.** The ability of heparin to inhibit spike cleavage at the S1/S2 site by furin was evaluated by using the colorimetric assay "CoviDrop$^{TM}$ SARS-Cov-2 targeted proprotein convertase inhibitor screening fast kit" (Epigentek, County Blvd, Farmingdale NY). Briefly, a 2.0 kDa SARS-CoV-2 spike fragment containing the intact S1/S2 $_{681}$RRAR$_{684}$ HBD is tagged with polyhistidine and biotin at its N-terminus and C-terminus, respectively. The spike fragment is then immobilized onto microplate wells through histidine/Ni-NTA. The cleavage of the substrate at the S1/S2 site removes the C-terminal S2 fragment of spike that is linked to biotin, causing a decrease of the signal generated by avidin/biotin binding that is detected by an appropriate colorimetric reaction (recorded by the absorbance in a microplate spectrophotometer at a wavelength of 450 nm. See Tab. S8-S9 for values of single measurements). Furin cleavage inhibition blocks the reduction of the signal, consequently the extent of spike cleavage is inversely proportional to the signal intensity. The assay was performed according to the manufacturer's instructions (https://www.epigentek.com/catalog/covidrop-sars-cov-targeted-proprotein-convertase-inhibitor-screening-fast-kit-p-84596.html).

## Acknowledgments

We acknowledge PRACE for awarding us access to Marconi100 based in Italy at CINECA (Project COVID19-54). The technical support of Alessandro Grottesi from CINECA (Italy) and Filippo Spiga from NVIDIA is gratefully acknowledged. G.P., S.R. and R.C.W. thank the Klaus Tschira Foundation and the Deutsche Forschungsgemeinschaft (DFG, German Research Foundation - Project number: 458623378 to R.C.W.) for support. G.P.

was supported by Erasmus+, an EMBO short-term fellowship (STF_8594) and The Guido Berlucchi foundation young researchers mobility program.

**Author Contributions.** G.P. Conceptualization, Methodology, Validation, Formal Analysis, Investigation - Simulations, Writing - Original Draft, Visualization, Funding acquisition. S.R. Software, Data Curation. P.O. Resources (glycosaminoglycans). C.U. Investigation - Experiments. M.R. Conceptualization, Resources, Writing - Review & Editing. R.W. Conceptualization, Resources, Writing - Review & Editing, Supervision, Funding acquisition.

**Competing Interest Statement.** The authors declare no competing interest.

**Additional information**
**Preprint server.** aXriv - https://arxiv.org/abs/2103.07722
**Data availability statement.** All simulation trajectories are available on the BioExcel COVID-19 platform: https://bioexcel-cv19.bsc.es/#/ with the identifiers from MCV1900217 to MCV1900236.
**Supplementary information**. Supplementary figures and tables, captions for movies M1-M3 and associated references (PDF). Enlargements of IFP-model02, IFP-model03 and IFP-model05 (tif). Movies S1, S2 andS3 (MP4).

# Supplementary Information for:

Three-fold mechanism of inhibition of SARS-CoV-2 infection by the interaction of the spike glycoprotein with heparin.

Giulia Paiardi*[1,2], Stefan Richter[1], Pasqua Oreste[3], Chiara Urbinati[2], Marco Rusnati[2], Rebecca C. Wade*[1,4].

[1] Molecular and Cellular Modeling Group, Heidelberg Institute for Theoretical Studies (HITS), 69118 Heidelberg, Germany,
[2] Macromolecular Interaction Analysis Unit, Section of Experimental Oncology and Immunology, Department of Molecular and Translational Medicine, 25123 Brescia, Italy,
[3] Glycores 2000 Srl, Milan, Italy.
[4] Zentrum für Molekulare Biologie (ZMBH), DKFZ-ZMBH Alliance and Interdisciplinary Center for Scientific Computing (IWR), Heidelberg University, 69120 Heidelberg, Germany,

*Corresponding authors: Giulia Paiardi, Rebecca C. Wade

**Email:** giulia.paiardi@h-its.org, Rebecca.wade@h-its.org

**Preprint server.** aXriv - https://arxiv.org/abs/2103.07722

**Data sharing. Data availability statement.** All simulation trajectories are available on the BioExcel COVID-19 platform: https://bioexcel-cv19.bsc.es/#/ with the identifiers from MCV1900217 to MCV1900336.

**This PDF file includes:**
1. Supplementary Figures: S1 to S9 --------------------------------------------------- 2-10
2. Supplementary Tables: S1 to S10 -------------------------------------------------- 11-22
3. Supplementary Movies: S1 to S3-------------------------------------------------- 23
4. Supplementary References ------------------------------------------------- 23

**Other supplementary materials for this manuscript include the following:**

IFP-model02, IFPmodel03, IFP-model05 (enlargement of single pictures of Tab.S6)
Movies S1-S3

In the following contents, a simplified nomenclature is used for the simulated systems in which they are referred to as Model01-05 and contain the following components:

| Model01 | Spike in closed conformation |
| Model02 | Closed spike + 1 heparin chain |
| Model03 | Closed spike + 3 heparin chains |
| Model04 | Open spike |
| Model05 | Open spike + 3 heparin chains |

# 1. Supplementary Figures

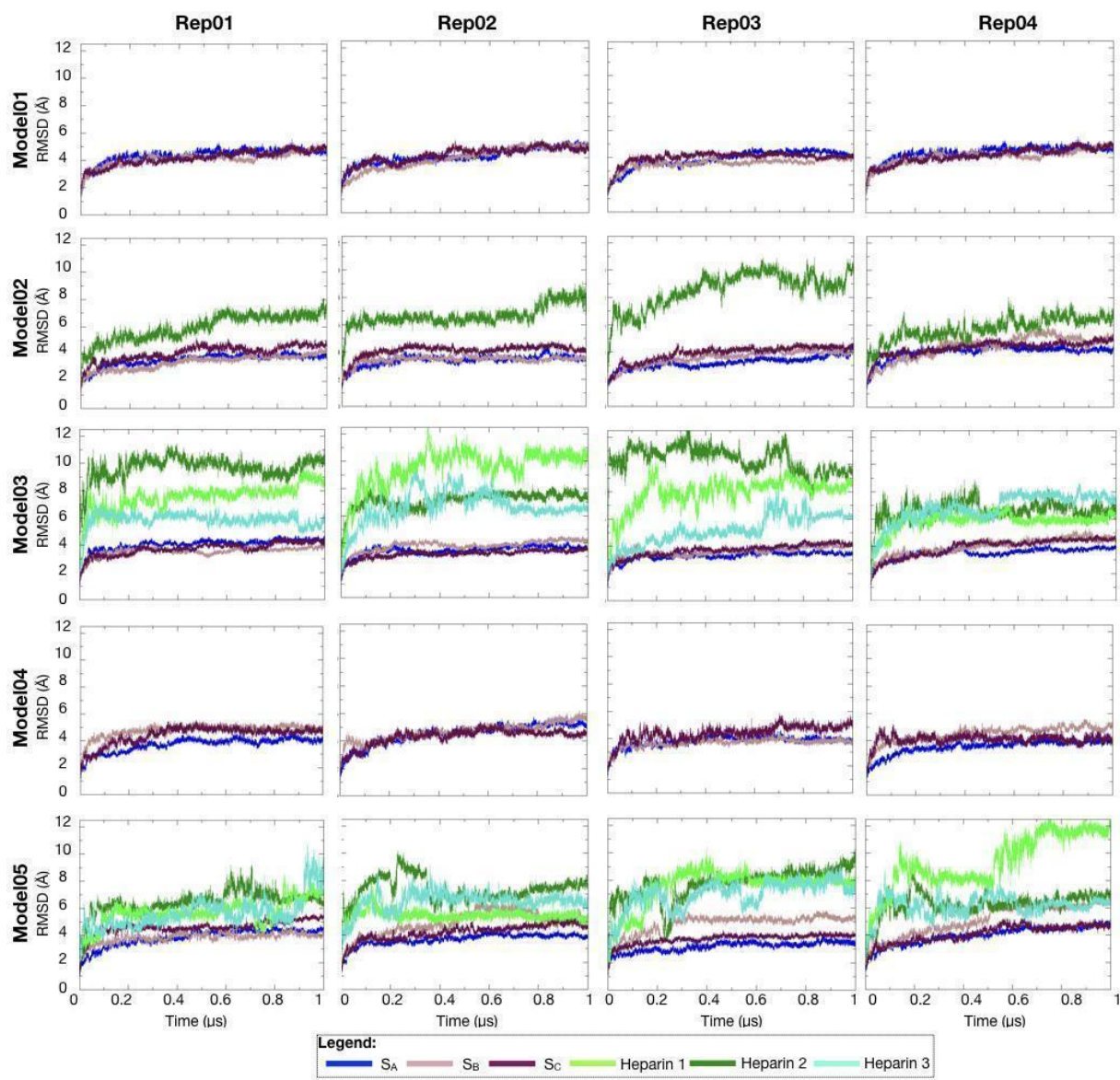

**Fig. S1 Structural convergence of the simulated systems.** Root mean square deviation (RMSD (Å)) versus time (µs) for the four replica MD trajectories of the five simulated systems. The RMSD values for the individual subunits - $S_A$, $S_B$ and $S_C$ - were calculated for the C-alpha atoms of residues 51-1063 and are shown in blue, pink and magenta, respectively. The RMSD values of the three heparin chains were calculated for all the C, N and O atoms in all monosaccharides and are colored green, light green and turquoise.

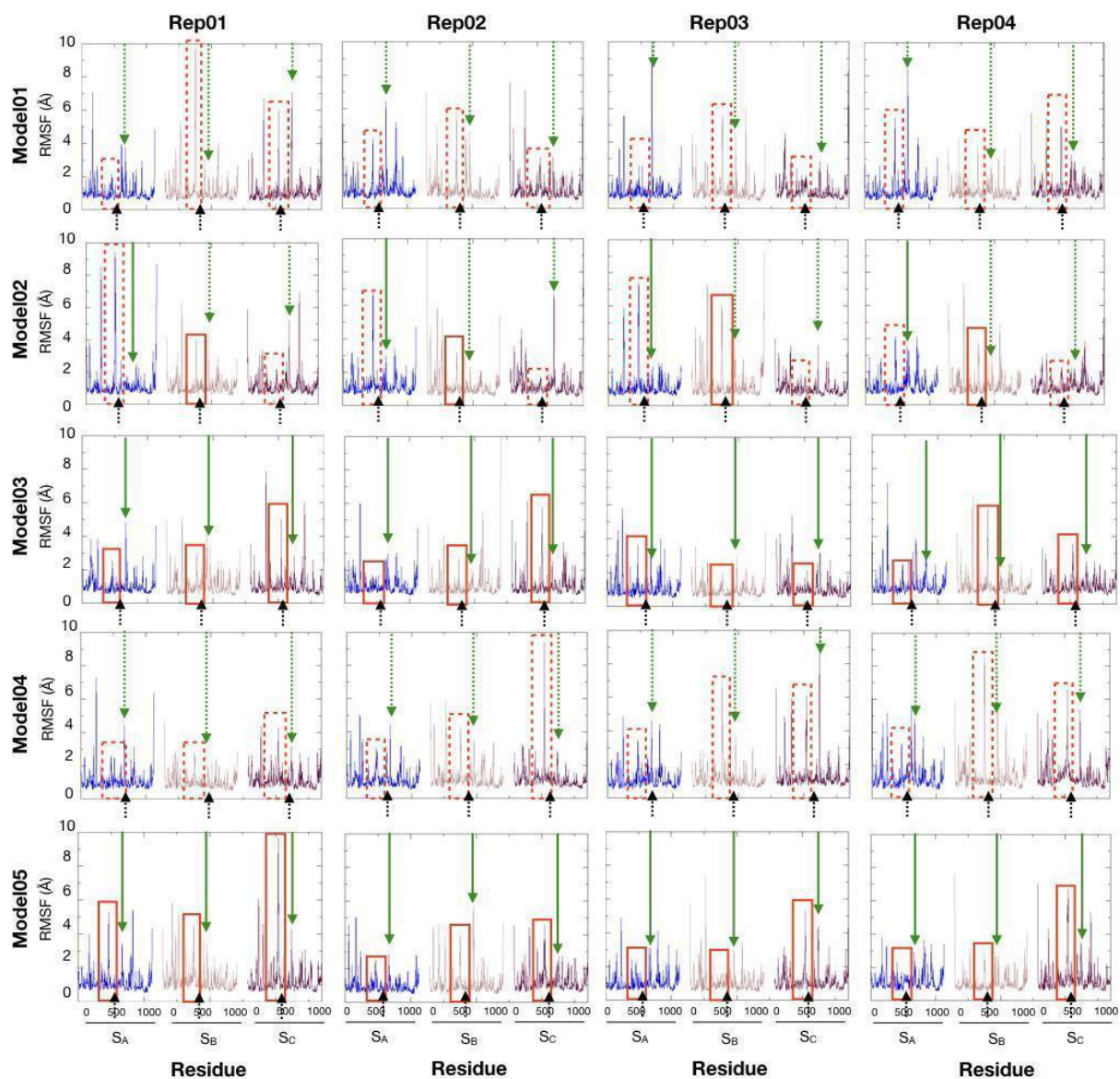

**Fig. S2 Conformational fluctuations of the RBD and the S1/S2 site.** Root mean square fluctuation (RMSF) (Å) versus residue number of the three subunits of the spike homotrimer for the four replica MD trajectories of the five simulated systems. The RMSF values for the individual subunits - $S_A$, $S_B$ and $S_C$ - were calculated for the C-alpha atoms of residues 1-3697 and are shown in blue, pink and magenta, respectively. The red boxes represent the RBD (residues 330-530) of each subunit, the black arrows highlight the hinge regions (residue 527-529), and the green arrows point to the S1/S2 residues (residues 682-685). For both the boxes and the arrows, the dashed and continuous lines represent the regions not interacting and interacting with heparin, respectively.

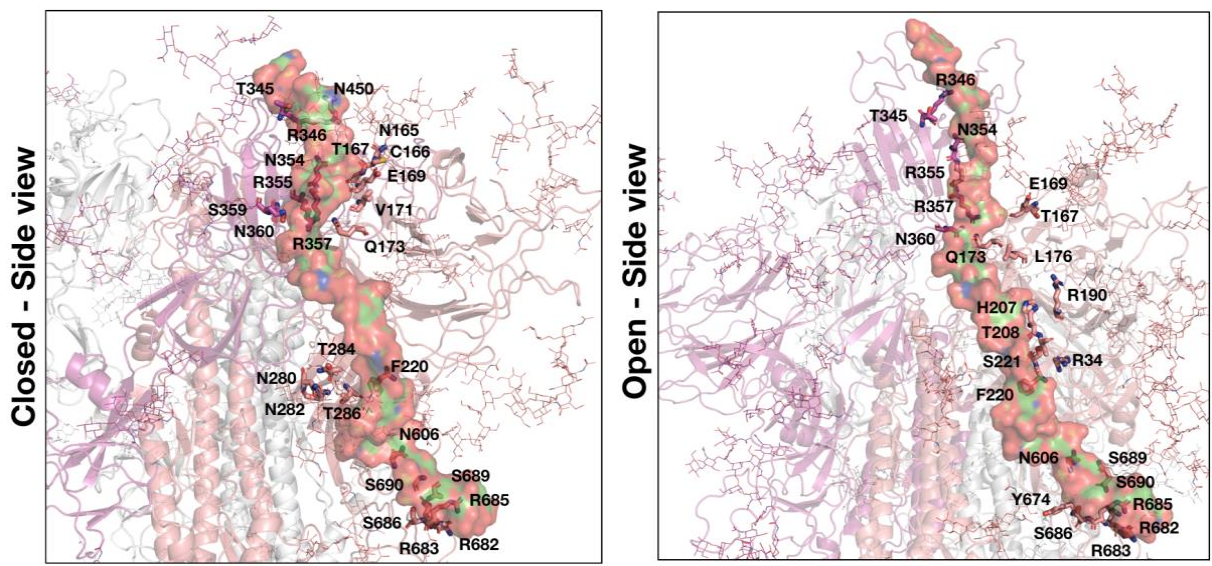

**Fig. S3 Heparin-spike hydrogen-bonding interactions.** Last snapshots of spike models in closed (left) and open (right) conformations with 3 heparins bound showed the H-bond interactions of heparin with spike subunits $S_B$ and $S_C$ as in close-ups of the side views in Figure. 1. The labelled residues shown in sticks and colored as the subunits were involved in H-bond interactions with heparin for more than half of the simulations of the spike in the closed (with 1 and 3 heparin chains) or open states. The $S_A$, $S_B$ and $S_C$ subunits are colored white, pink and magenta, respectively, and are shown in cartoon representation with the covalently attached N-glycans in line representation. Heparin is shown as a surface coloured by elements in green, red, yellow and blue for carbons, oxygens, sulphates and nitrogens, respectively.

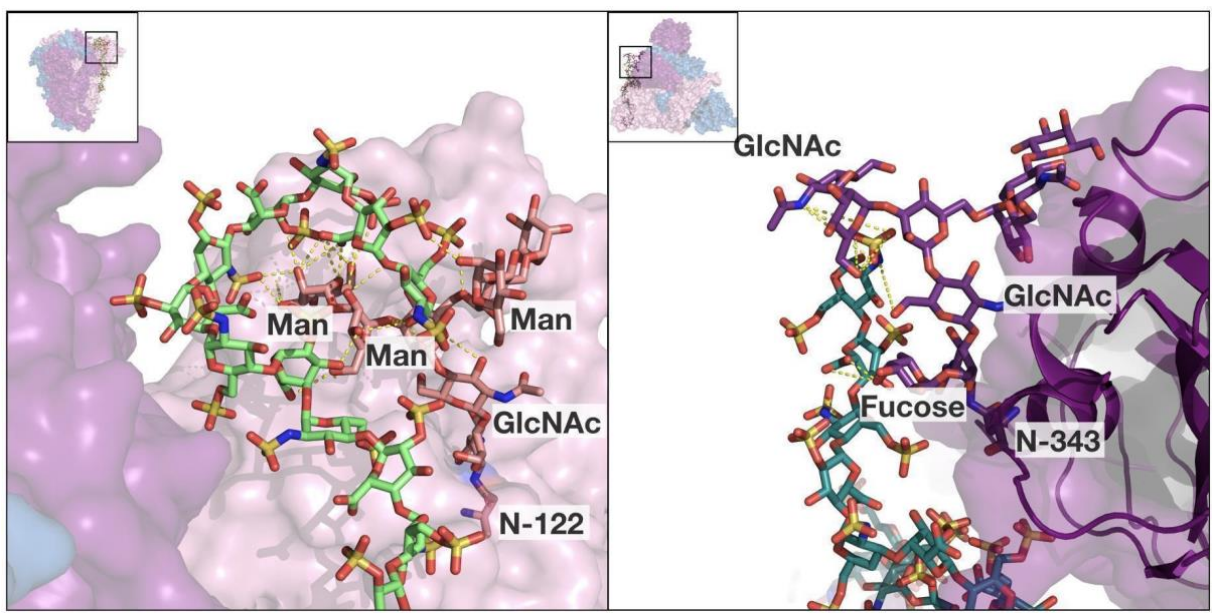

**Fig. S4 Spike N-glycan interactions with heparin.** N-glycans can modulate binding of the spike to heparin by transiently displacing heparin from the spike surface (see Movies S1-S3). The heparin chains in both closed and open models of spike were modelled to interact with the K444 and N448 residues, that are in close proximity to the ACE2 binding residues. Visual inspection of the trajectories and the H-bond analysis showed that N-glycans can aspecifically and transiently cause the detachment of a limited portion of the heparin chains from the spike. We identified two N-glycans with key roles in this mechanism: the N-glycan attached to N-122 (in the N-terminal domain) acts only in the closed conformation whereas the N-glycan attached to N-343 (in the RBD) is mainly responsible for detachment in both the closed and the open conformations. The figure shows the interactions between the N-122 (left) and N-343 (right) N-glycans with heparin in closed and open models, respectively. The region shown in the spike is indicated by the squares in the insets. The $S_B$ and $S_C$ subunits are shown as translucent surfaces and cartoons in pink and magenta, respectively. N-122, N-343 and the N-glycans are shown in stick representation and colored by element with pink/magenta carbons. Heparin chains are shown in stick representation with carbons in light green (left) and cyan (right). Dashed lines indicate the H-bond interactions between the glycans and heparin. As SPR and circular dichroism spectroscopy experiments were done using unfractionated heparins (13.5-15kDa - ~48 monosaccharides), we cannot exclude that longer heparin chains could maintain the interaction with these residues (1). Our simulations suggest that N-glycans may exert a shielding effect that results in a non-specific, partial and transient detachment of heparin from the spike.

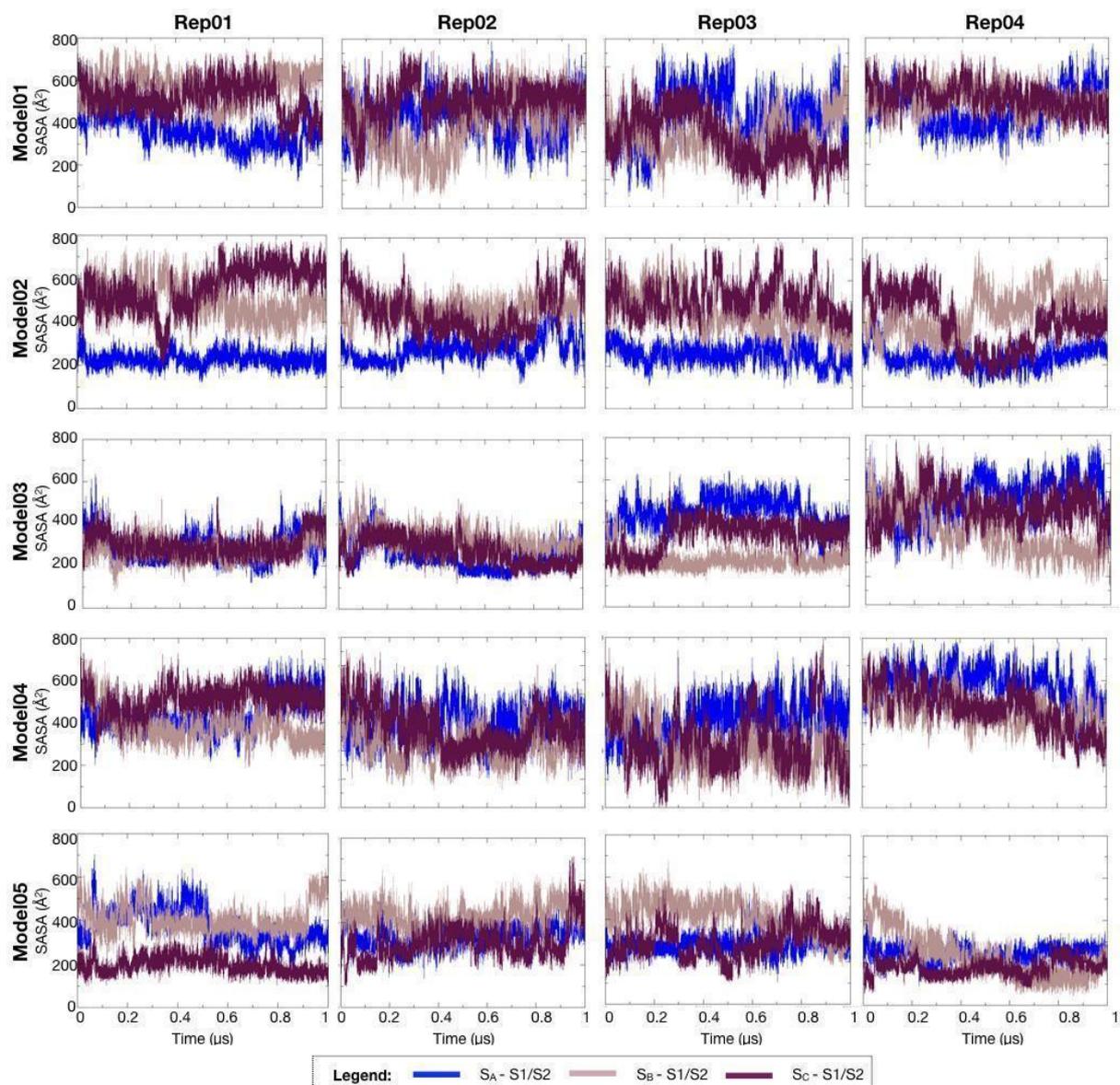

**Fig. S5 Exposure of the S1/S2 site during the MD simulations.** Solvent Accessible Surface Area (SASA) (Å$^2$) of the S1/S2 site versus time (µs) for the four replica MD trajectories of the five simulated systems. The SASA of the S1/S2 site (residues 682-685) in each spike subunit - S$_A$, S$_B$ and S$_C$ - is shown in blue, pink and magenta, respectively, and was computed using CPPTRAJ (2).

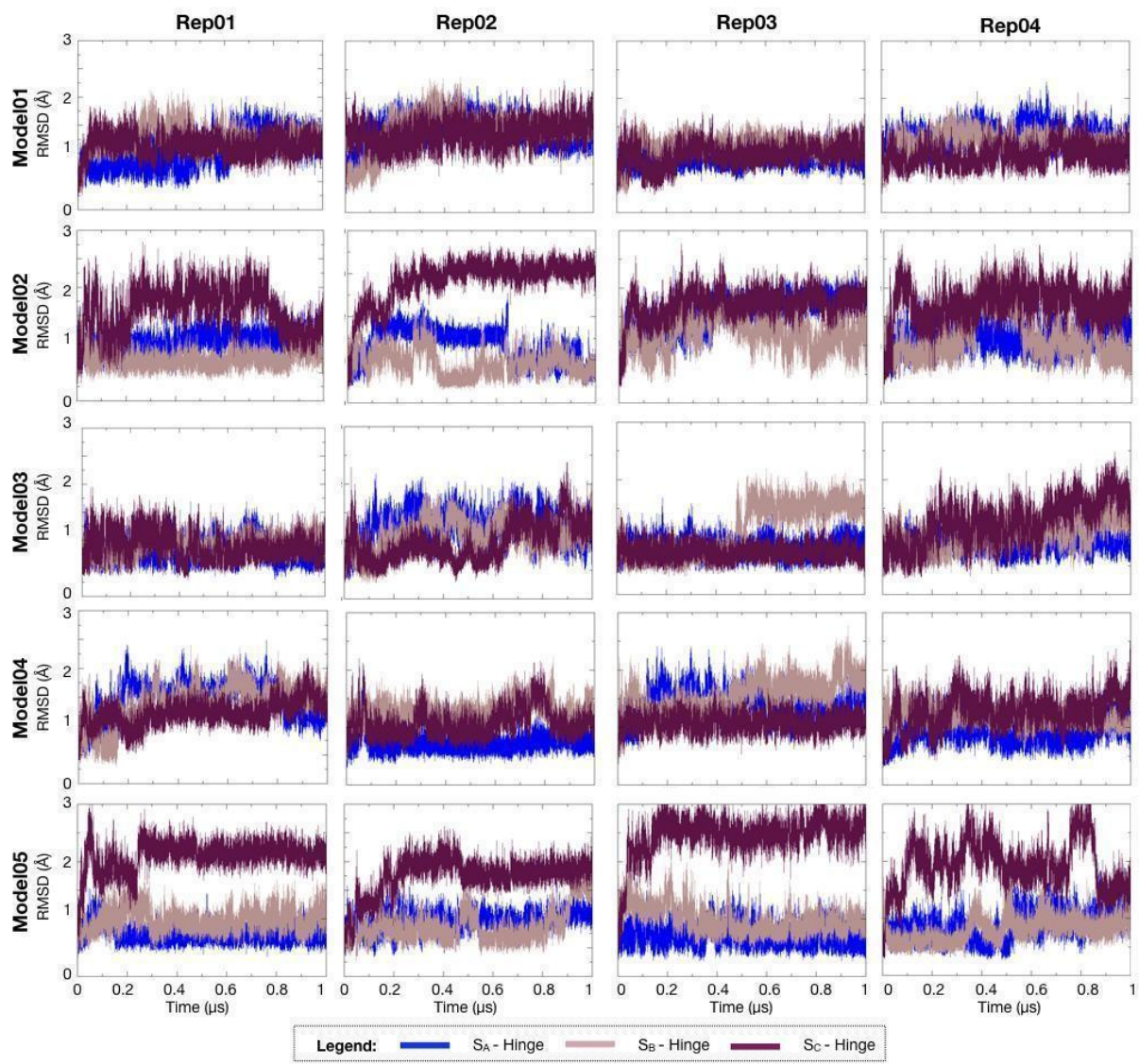

**Fig. S6 Conformational changes of the hinge region during the simulations.** Root mean square deviation (Å)) of the hinge region versus time (μs) for the four replica MD trajectories of the five simulated systems. The RMSD values were calculated for the C-alpha atoms of residues 527-529 and plotted for the individual subunits - $S_A$, $S_B$ and $S_C$ - in blue, pink and magenta, respectively.

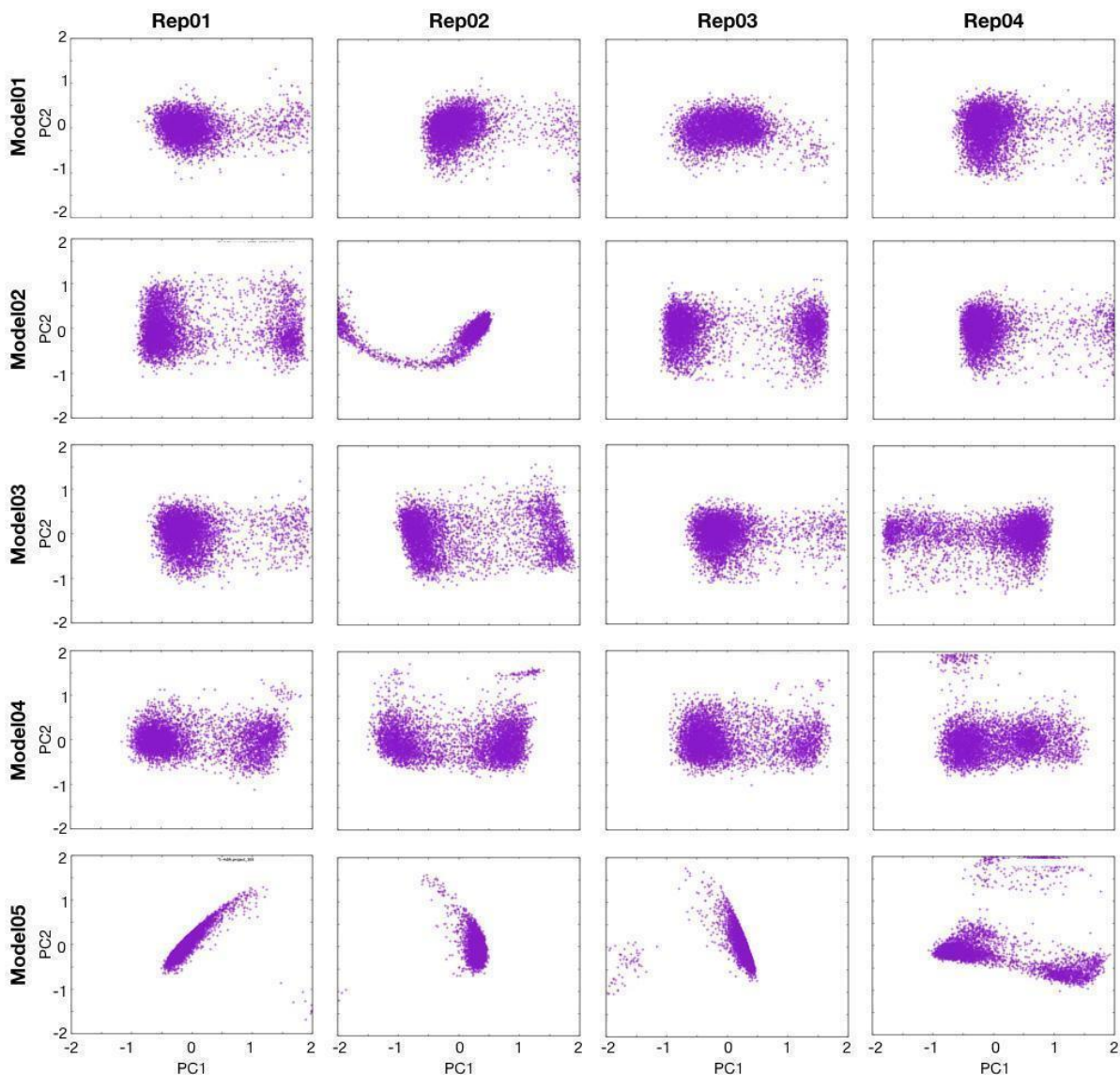

**Fig. S7 Effects of heparin binding on the motion of the RBD-hinge region.** Results of dihedral Principal Component Analysis (dPCA) calculated for the Phi/Psi angles of the RBD-hinge region (residues 527-529) of the $S_C$ subunit for the four replica MD trajectories of the five simulated systems. PC1 and PC2 are plotted on the x and y axes, respectively. Note that the second replica of model 02 differs from the others because of the interaction between heparin and the N-glycan at N122 and the fourth replica of model 05 differs from the others for this system because of the interaction between heparin and the N-glycan at N343 in this trajectory.

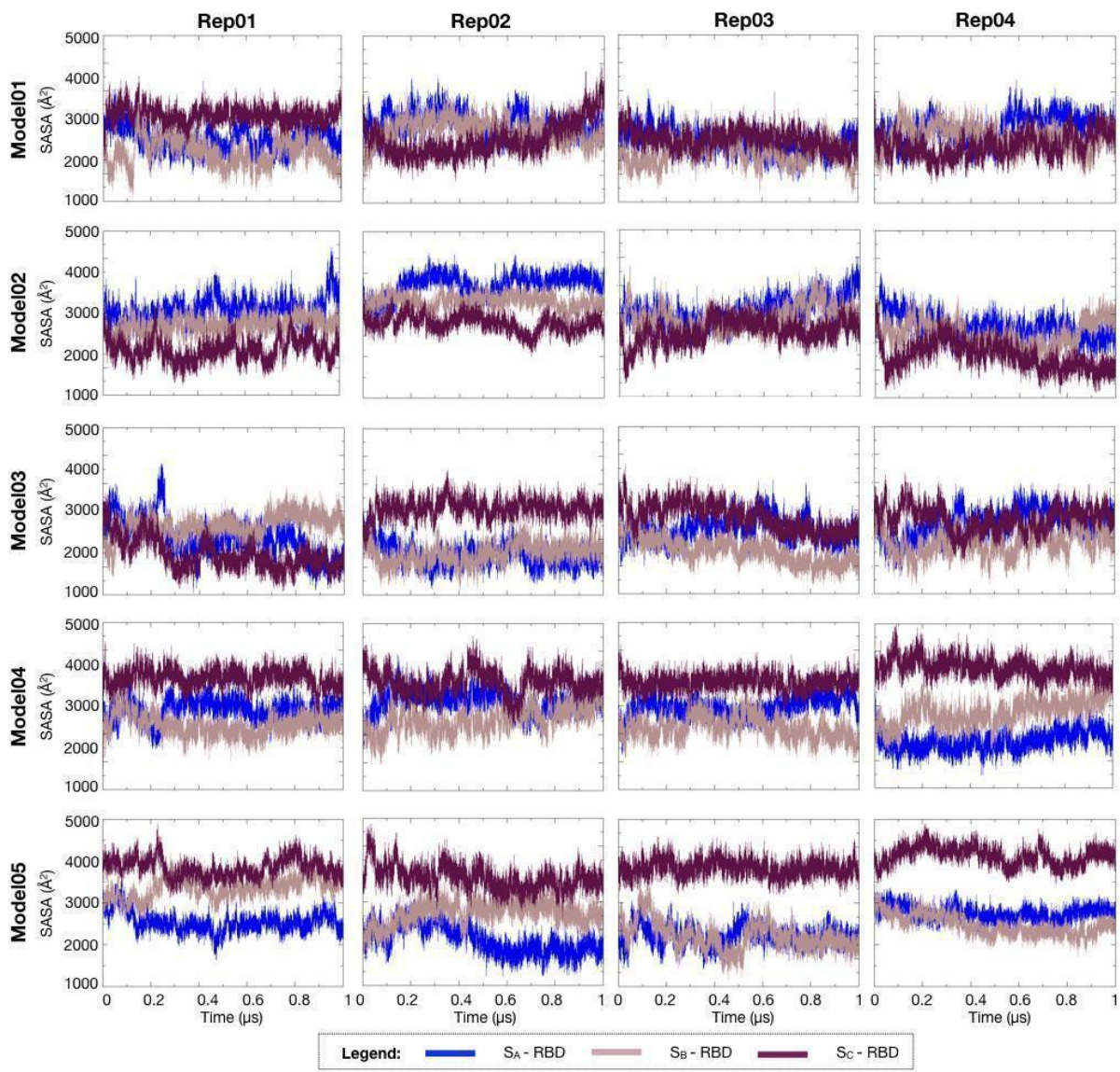

**Fig. S8 Exposure of the receptor binding motif (RBm) during the MD simulations.** Solvent Accessible Surface Area (SASA) (Å$^2$) of the RBm versus time (μs) for the four replica MD trajectories of the five simulated systems. The SASA is shown for each subunit - $S_A$, $S_B$ and $S_C$ - in blue, pink and magenta, respectively, and was calculated using CPPTRAJ (2).

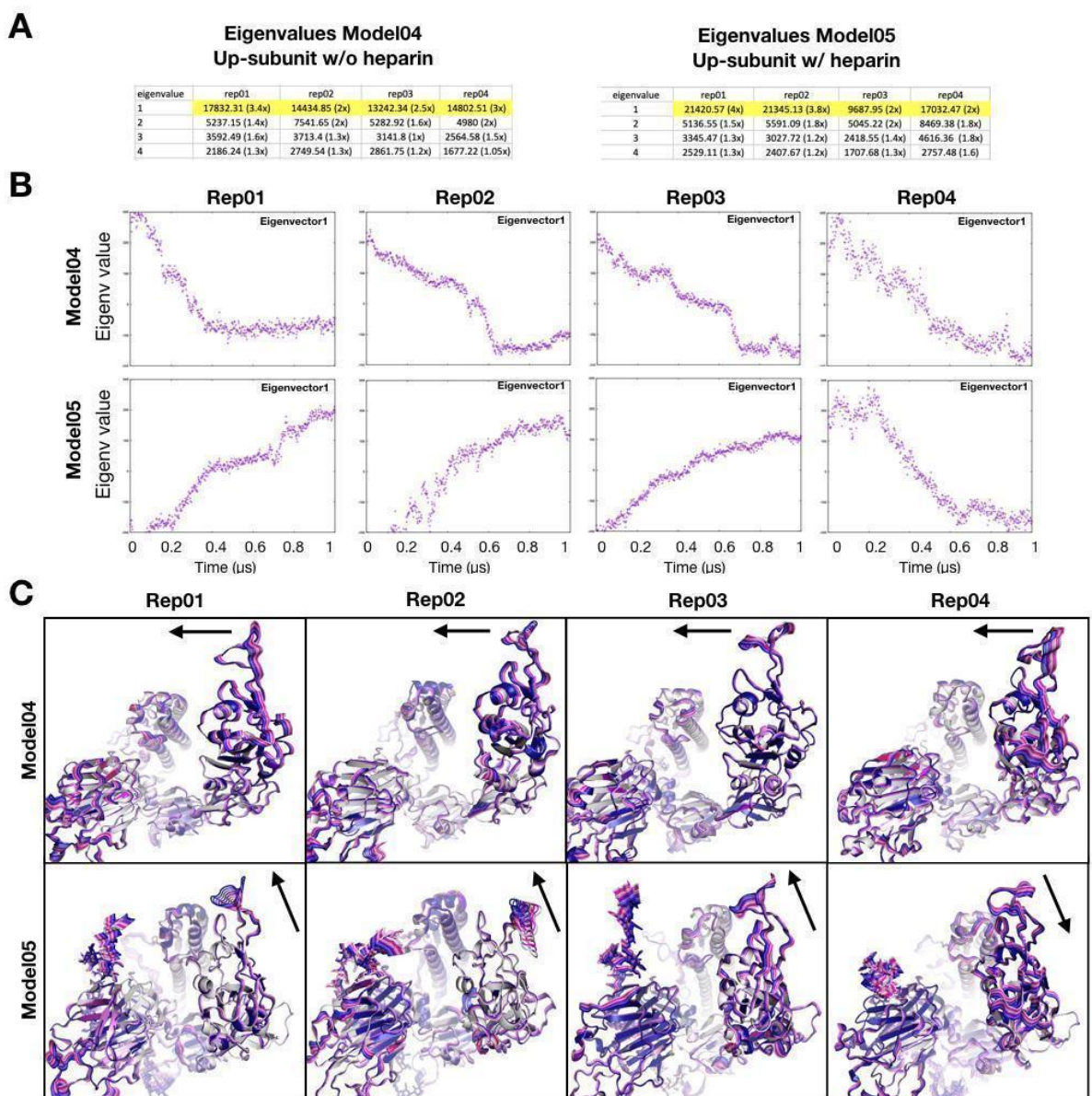

**Fig. S9 Differences in the spike RBD opening motion in the presence of heparin.** Essential Dynamics (ED) analysis on the $S_C$ subunit for the four replica MD trajectories of model04 and model05. (A) Values of the first four eigenvalues for the up-subunit of Model04 and Model05. (B) Plot of the first normalized eigenvector along the trajectory. (C) Plot of the direction of motion of the first eigenvector on the RBD of the $S_C$ subunit which can be seen to differ between Model04 and Model05. Note that the direction for the fourth replica of Model05 differs from the other replicas because of the interaction between heparin and the N-glycan at N343 in this trajectory.

## 3. Supplementary Tables

**Tab. S1 Average root mean square fluctuation (RMSF) (Å) calculated for the C-alpha atoms of the RBD (residues 330-530).** The average fluctuation is calculated based on the average fluctuation of the single residues in the RBD. In model02, the single heparin chains directly bind to the RBD of monomer $S_C$. The grey boxes represent the subunits interacting with heparin.

|  |  | Subunit SA | | | | Subunit SB | | | | Subunit SC | | | | AVG [Å] | ST.DEV. [Å] |
|---|---|---|---|---|---|---|---|---|---|---|---|---|---|---|---|
|  |  | Rep 01 | Rep 02 | Rep 03 | Rep 04 | Rep 01 | Rep 02 | Rep 03 | Rep 04 | Rep 01 | Rep 02 | Rep 03 | Rep 04 | | |
| **Average Fluctuation RBD residues [Å]** | Model 01 | 10.18 | 9.40 | 9.99 | 9.44 | 10.22 | 11.18 | 13.72 | 13.37 | 11.59 | 11.19 | 11.14 | 10.52 | 10.83 | 1.38 |
|  | Model 02 | 10.23 | 13.93 | 13.80 | 13.65 | 10.68 | 10.69 | 12.14 | 9.91 | 10.10 | 10.14 | 9.51 | 9.60 | 10.45 | 1.71 |
|  | Model 03 | 9.39 | 8.83 | 9.38 | 9.19 | 8.51 | 9.77 | 8.52 | 12.05 | 10.79 | 10.30 | 9.93 | 10.88 | 9.57 | 1.06 |
|  | Model 04 | 10.57 | 12.12 | 10.88 | 10.56 | 14.89 | 12.53 | 14.54 | 10.94 | 17.38 | 15.29 | 13.93 | 13.01 | 12.77 | 2.19 |
|  | Model 05 | 10.09 | 9.80 | 9.27 | 12.10 | 9.18 | 11.25 | 9.45 | 14.18 | 15.57 | 14.54 | 12.05 | 16.88 | 11.65 | 2.67 |

**Tab. S2 Average root mean square fluctuation (RMSF) (Å) calculated for the C-alpha atoms of the hinge region (residues 527-529).** The average fluctuation is calculated based on the average fluctuation of the single residues in the hinge region. In model02, the single heparin chain allosterically acts on the hinge region of monomer $S_C$. The grey boxes represent the subunits interacting with heparin.

|  |  | Subunit SA | | | | Subunit SB | | | | Subunit SC | | | | AVG. [Å] | ST.DEV. [Å] |
|---|---|---|---|---|---|---|---|---|---|---|---|---|---|---|---|
|  |  | Rep 01 | Rep 02 | Rep 03 | Rep 04 | Rep 01 | Rep 02 | Rep 03 | Rep 04 | Rep 01 | Rep 02 | Rep 03 | Rep 04 | | |
| **Average Fluctuation hinge region [Å]** | Model01 | 1.07 | 0.8 | 1.0 | 1.0 | 1.0 | 1.0 | 1.0 | 1.0 | 1.0 | 1.0 | 1.0 | 1.0 | 1.0 | 0.06 |
|  | Model02 | 1.14 | 1.2 | 1.07 | 0.8 | 1.12 | 1.1 | 0.7 | 1.0 | 1.3 | 1.15 | 1.15 | 1.3 | 1.13 | 0.18 |
|  | Model03 | 0.94 | 1.0 | 0.94 | 1.0 | 0.9 | 1.07 | 0.7 | 1.07 | 0.97 | 1.07 | 1.04 | 1.0 | 1.0 | 0.10 |
|  | Model04 | 1.4 | 1.1 | 1.34 | 1.67 | 1.24 | 1.3 | 1.3 | 1.24 | 1.5 | 1.5 | 1.27 | 1.27 | 1.3 | 0.15 |
|  | Model05 | 1.5 | 0.7 | 0.7 | 1.04 | 0.8 | 1.0 | 0.7 | 1.34 | 1.4 | 1.04 | 1.07 | 0.97 | 1.02 | 0.28 |

**Tab. S3 Average root mean square fluctuation (Å) calculated for the C-alpha atoms of the S1/S2 site (residues 682-685).** The average fluctuation is calculated based on the fluctuation of the single residues in the S1/S2 site. In model02, the single heparin chain binds the S1/S2 site of monomer $S_A$. The grey boxes represent the subunits interacting with heparin.

| | | Subunit SA | | | | Subunit SB | | | | Subunit SC | | | | AVG [Å] | ST.DEV. [Å] |
|---|---|---|---|---|---|---|---|---|---|---|---|---|---|---|---|
| | | Rep 01 | Rep 02 | Rep 03 | Rep 04 | Rep 01 | Rep 02 | Rep 03 | Rep 04 | Rep 01 | Rep 02 | Rep 03 | Rep 04 | | |
| **Average Fluctuation S1/S2 residues [Å]** | Model01 | 6.2 | 6.25 | 9.25 | 2.13 | 2.13 | 3.88 | 2.88 | 1.75 | 2 | 2.63 | 1.38 | 4.65 | 2.75 | 2.4 |
| | Model02 | 2.87 | 2.08 | 1.55 | 1.2 | 1.2 | 1.78 | 2 | 1.6 | 1.5 | 6 | 2.93 | 4.18 | 2.04 | 1.38 |
| | Model03 | 1.78 | 3.18 | 3 | 3.25 | 3.25 | 2.25 | 1.08 | 1.05 | 1.3 | 2.08 | 2.5 | 1.78 | 1.93 | 0.78 |
| | Model04 | 3.9 | 3.05 | 3.18 | 3.43 | 3.43 | 3.25 | 3 | 1.58 | 1.55 | 2.45 | 7.7 | 2.35 | 3.11 | 1.71 |
| | Model05 | 2.08 | 1.85 | 2.13 | 2.75 | 2.75 | 5.15 | 2.33 | 2.65 | 2.75 | 1.98 | 3.5 | 3.73 | 2.49 | 0.97 |

**Tab. S4 Analysis of hydrogen bonds (H-bonds) between the spike and the heparin chains in Model03 for the four replica MD trajectories using CPPTRAJ (2).** Each heparin chain interacts with two adjacent subunits, $S_A$-$S_B$, $S_B$-$S_C$ or $S_C$-$S_A$, and the spike residues involved in the interaction are listed separately as residues of the first subunit and of the second subunit. The yellow boxes represent the H-bonds that were stable in more than half the simulations with an occupancy > 50% in the single trajectory considering the 3 heparin chains on each subunit. The H-bonds for Model02 (data not shown) are not reported because they are very similar to these results.

| Spike Residues | Rep 01 | | | Rep 02 | | | Rep 03 | | | Rep 04 | | | SUM |
|---|---|---|---|---|---|---|---|---|---|---|---|---|---|
| First subunit | Heparin on $S_A$-$S_B$ subunits | Heparin on $S_B$-$S_C$ subunits | Heparin on $S_C$-$S_A$ subunits | Heparin on $S_A$-$S_B$ subunits | Heparin on $S_B$-$S_C$ subunits | Heparin on $S_C$-$S_A$ subunits | Heparin on $S_A$-$S_B$ subunits | Heparin on $S_B$-$S_C$ subunits | Heparin on $S_C$-$S_A$ subunits | Heparin on $S_A$-$S_B$ subunits | Heparin on $S_B$-$S_C$ subunits | Heparin on $S_C$-$S_A$ subunits | |
| ARG_34 | | | 1 | | | | | | 1 | | | | 2 |
| TYR_38 | | 1 | | 1 | | 1 | | | | | 1 | | 4 |
| LYS_41 | 1 | | | | | | 1 | | | | | 1 | 3 |
| ARG_158 | | | | | 1 | | | | | | | | 1 |
| SER_162 | | | | | 1 | | | | | | 1 | | 2 |
| ASN_164 | | | | | 1 | | | | | | 1 | | 2 |
| ASN_165 | | | 1 | 1 | 1 | | | | | | | | 3 |
| CYX_166 | | | | 1 | | 1 | | | | | | | 2 |
| THR_167 | | 1 | | | | 1 | | 1 | | | | 1 | 4 |
| GLU_169 | | 1 | | 1 | | | | | | 1 | | 1 | 4 |
| TYR_170 | | | | 1 | 1 | 1 | | | | | | | 3 |
| VAL_171 | 1 | | | 1 | 1 | | | | | | 1 | | 4 |
| SER_172 | | | | | | 1 | | | | | 1 | | 2 |
| GLN_173 | 1 | 1 | 1 | 1 | | 1 | | 1 | 1 | | 1 | | 8 |
| LEU_176 | | | 1 | | | | | 1 | | | | 1 | 3 |
| ASN_180 | 1 | | | | | 1 | | | | | | | 2 |
| THR_186 | | | | | 1 | | | | | | | | 1 |
| ARG_190 | | 1 | | | | | | 1 | | | | 1 | 3 |
| LYS_206 | | | | | | | | 1 | | | | | 1 |
| HIE_207 | | 1 | | | | | | 1 | | | 1 | 1 | 4 |
| THR_208 | | 1 | | | | | | 1 | | | 1 | 1 | 4 |
| GLN_211 | | | | | | | | | | | | 1 | 1 |
| PHE_220 | 1 | 1 | 1 | 1 | | | 1 | | 1 | 1 | 1 | | 8 |
| SER_221 | | 1 | 1 | | | | | 1 | | | 1 | | 4 |
| ASP_228 | | | 1 | | | | | | | | | | 1 |
| LYS_278 | 1 | | | 1 | 1 | | 1 | | | | 1 | | 5 |
| ASN_280 | 1 | 1 | | | | | | | | | | | 2 |
| THR_284 | 1 | 1 | | | | | | | | | | | 2 |
| THR_286 | 1 | 1 | | | 1 | | 1 | | | 1 | 1 | | 6 |
| ASP_287 | 1 | 1 | | 1 | 1 | | 1 | 1 | | 1 | | | 7 |
| LYS_300 | | | | | | | | 1 | 1 | | | | 2 |
| THR_308 | 1 | | | 1 | 1 | | 1 | | | | | | 4 |
| THR_602 | 1 | | | 1 | 1 | | 1 | | | | 1 | | 5 |
| ASN_603 | 1 | | | 1 | 1 | | 1 | 1 | | 1 | 1 | | 7 |
| THR_604 | | | | 1 | | | 1 | | | | | | 2 |
| SER_605 | | | | 1 | | | | | | | | 1 | 2 |
| ASN_606 | 1 | 1 | 1 | 1 | 1 | 1 | 1 | 1 | 1 | 1 | | 1 | 11 |
| GLN_607 | | 1 | 1 | | | | | 1 | | | | | 3 |
| TYR_674 | | | | 1 | 1 | | | | | | | 1 | 3 |
| GLN_677 | | 1 | | | | | | | | | | | 1 |
| SER_690 | 1 | | | | | | | | | | | | 1 |
| ARG_682 | 1 | | 1 | 1 | | 1 | 1 | | 1 | 1 | | 1 | 8 |
| ARG_683 | 1 | 1 | 1 | 1 | | | | | 1 | 1 | 1 | 1 | 8 |
| ALA_684 | | 1 | | | | | | | | | 1 | | 2 |
| ARG_685 | 1 | 1 | 1 | 1 | 1 | 1 | 1 | 1 | 1 | 1 | 1 | 1 | 12 |
| SER_686 | 1 | 1 | 1 | 1 | | | 1 | 1 | | | 1 | 1 | 7 |
| VAL_687 | | 1 | | | | | 1 | 1 | | | 1 | | 4 |
| SER_689 | | 1 | 1 | 1 | | | | | | | | | 3 |
| GLN_690 | | 1 | | | | 1 | | | | | | 1 | 3 |
| SER_940 | | | | 1 | | | | | | | | | 1 |
| **Second subunit** | | | | | | | | | | | | | |
| ASN_334 | | | | 1 | | | | 1 | | | | | 2 |
| LEU_335 | | | | 1 | | 1 | | | | | | | 2 |
| PHE 338 | | | | | | | | | | | 1 | | 1 |
| GLY 3239 | | | | | | | | | | | 1 | | 1 |
| GLU_340 | | | | 1 | | | | | | | | | 1 |

| Residue | C1 | C2 | C3 | C4 | C5 | C6 | C7 | C8 | C9 | C10 | C11 | C12 | Total |
|---|---|---|---|---|---|---|---|---|---|---|---|---|---|
| ASN_343 |  | 1 |  |  |  |  |  | 1 | 1 |  |  |  | 3 |
| THR 345 | 1 | 1 |  |  |  | 1 |  | 1 | 1 | 1 |  | 1 | 7 |
| ARG_346 | 1 | 1 | 1 | 1 | 1 | 1 |  | 1 | 1 | 1 | 1 | 1 | 11 |
| ASN_353 |  |  | 1 |  | 1 | 1 |  |  |  |  |  | 1 | 4 |
| ARG_355 |  |  |  |  |  |  |  |  |  | 1 |  | 1 | 2 |
| LYS_356 |  | 1 |  | 1 |  |  |  | 1 |  | 1 |  |  | 4 |
| ARG_357 |  | 1 | 1 | 1 | 1 | 1 | 1 | 1 | 1 | 1 | 1 | 1 | 11 |
| SER_359 | 1 | 1 |  |  | 1 |  |  | 1 | 1 |  |  |  | 5 |
| ASN_360 |  |  | 1 |  | 1 |  | 1 |  |  |  | 1 | 1 | 5 |
| ASP_364 |  |  |  |  |  |  |  |  |  | 1 |  |  | 1 |
| TYR_365 |  |  |  |  |  |  |  |  |  | 1 |  |  | 1 |
| SER_366 |  |  |  |  |  |  |  |  |  | 1 |  |  | 1 |
| VAL_367 |  |  |  |  |  |  |  |  |  | 1 |  |  | 1 |
| ASN_394 |  |  |  |  |  |  | 1 |  |  |  |  |  | 1 |
| TYR_396 |  |  |  |  |  |  |  |  |  |  | 1 |  | 1 |
| LYS_444 |  |  |  |  | 1 |  |  |  |  |  |  |  | 1 |
| VAL_445 |  |  |  |  | 1 | 1 |  |  |  |  |  |  | 2 |
| TYR_449 |  |  |  |  | 1 | 1 |  |  |  |  |  |  | 2 |
| ASN_450 |  |  |  |  | 1 |  |  |  |  |  | 1 |  | 2 |
| ARG_466 |  |  |  |  |  |  |  |  |  |  | 1 |  | 1 |
| ARG_509 | 1 |  |  |  |  |  |  |  |  |  |  |  | 1 |
| LYS_529 |  |  |  |  |  |  | 1 |  |  |  |  |  | 1 |
| GLN_563 | 1 |  |  |  |  |  | 1 |  |  |  |  |  | 1 |
| GLN_564 | 1 |  |  |  |  |  |  |  |  |  |  |  | 1 |
| ARG_577 | 1 |  |  | 1 |  |  | 1 |  |  |  |  |  | 3 |
| THR_581 | 1 |  |  |  |  |  |  |  |  |  |  |  | 1 |

**Tab. S5 Hydrogen Bond Analysis (H-bonds) of Model05 between S glycoprotein and the heparin chains.**
The data are collected for the four replicas using the H-bond analysis implemented in CPPTRAJ (2). Each heparin chain interacts with two adjacent subunits, $S_A$-$S_B$, $S_B$-$S_C$ or $S_C$-$S_A$, and the spike residues involved in the interaction are listed separately as residues of the first subunit and of the second subunit. The yellow boxes represent the H-bonds that were stable in more than half the simulations with an occupancy > 50% in the single trajectory considering the 3 heparin chains on each subunit.

| Spike Residues | Rep 01 | | | Rep 02 | | | Rep 03 | | | Rep 04 | | | SUM |
|---|---|---|---|---|---|---|---|---|---|---|---|---|---|
| | Heparin on $S_A$-$S_B$ subunits | Heparin on $S_B$-$S_C$ subunits | Heparin on $S_C$-$S_A$ subunits | Heparin on $S_A$-$S_B$ subunits | Heparin on $S_B$-$S_C$ subunits | Heparin on $S_C$-$S_A$ subunits | Heparin on $S_A$-$S_B$ subunits | Heparin on $S_B$-$S_C$ subunits | Heparin on $S_C$-$S_A$ subunits | Heparin on $S_A$-$S_B$ subunits | Heparin on $S_B$-$S_C$ subunits | Heparin on $S_C$-$S_A$ subunits | |
| **First subunit** | | | | | | | | | | | | | |
| ARG_34 | 1 | | | | | | 1 | | 1 | 1 | | | 4 |
| ASN_125 | | | | | | | 1 | | | | | | 1 |
| LYS_129 | 1 | | | | | | 1 | | | | | | 2 |
| ARG_158 | | | | | | | 1 | | | | 1 | 1 | 3 |
| TYR_160 | | | | 1 | | | | | | | | | 1 |
| ALA_163 | | | | | | | | | | | | | 0 |
| ASN_164 | | 1 | | | 1 | | 1 | | | | | | 3 |
| ASN_165 | | | | 1 | | | 1 | | 1 | 1 | | | 4 |
| CYX_166 | 1 | | | | | | 1 | 1 | | 1 | 1 | | 5 |
| THR_167 | 1 | 1 | | | | | | 1 | | | 1 | | 4 |
| GLU_169 | | 1 | 1 | | 1 | | 1 | | | | 1 | 1 | 6 |
| TYR_170 | | | | | | | | | | | 1 | | 1 |
| VAL_171 | | 1 | 1 | | | | | 1 | | | | 1 | 4 |
| SER_172 | | | 1 | | | | | 1 | 1 | | | | 3 |
| GLN_173 | | | 1 | | 1 | 1 | | | 1 | 1 | | | 5 |
| LYS_187 | | | | | | | 1 | | | | | | 1 |
| LYS_206 | | | | | | | 1 | | | | | | 1 |
| HIE_207 | 1 | | | | | | 1 | | | | | | 2 |
| THR_208 | | | | | | | | | | | | | 0 |
| ASN_211 | | 1 | | | | | | | | | | | 1 |
| GLN_218 | | 1 | | | | | | | | | | | 1 |
| PHE_220 | | | 1 | 1 | 1 | | 1 | | 1 | | | | 5 |
| SER_221 | | | 1 | | | | | | | | | | 1 |
| LYS_278 | | | 1 | | | | | | | | | | 1 |
| ASN_280 | | 1 | 1 | 1 | 1 | | 1 | 1 | 1 | | | 1 | 8 |
| ASN_282 | | 1 | 1 | 1 | | | | 1 | | | | 1 | 5 |
| THR_284 | | 1 | 1 | 1 | | | | 1 | 1 | | | 1 | 6 |
| THR_286 | | 1 | 1 | 1 | 1 | | 1 | 1 | 1 | | | | 7 |
| ASP_287 | | 1 | | | | | | | | | | | 1 |
| THR_307 | | | | | | | | | | | | 1 | 1 |
| THR_604 | | | | | | | 1 | | | | | | 1 |
| SER_605 | | | | | | | | 1 | | | | | 1 |
| ASN_606 | 1 | | 1 | | | | | 1 | | 1 | 1 | 1 | 6 |
| GLN_607 | 1 | | | | | | | | | | 1 | | 2 |
| TYR_674 | | 1 | | | | | | 1 | 1 | | | | 3 |
| ASN_679 | | | | | | | 1 | | | | | | 1 |
| SER_680 | | 1 | | | | | 1 | | | | | | 2 |
| ARG_682 | 1 | 1 | | 1 | 1 | 1 | 1 | | | | 1 | 1 | 8 |
| ARG_683 | 1 | 1 | 1 | 1 | 1 | 1 | 1 | 1 | | 1 | 1 | 1 | 11 |
| ALA_684 | 1 | | | | 1 | | | | | 1 | | | 3 |
| ARG_685 | 1 | 1 | 1 | 1 | 1 | 1 | 1 | 1 | 1 | 1 | 1 | 1 | 12 |
| SER_686 | | | 1 | 1 | 1 | | 1 | | | | 1 | 1 | 6 |
| VAL_687 | | | | | | | 1 | | | | | | 1 |
| ALA_688 | | | | | | | 1 | | | | | | 1 |
| SER_689 | 1 | | 1 | | | 1 | 1 | 1 | | | | | 5 |
| GLN_690 | | | 1 | | | 1 | 1 | 1 | | | | | 4 |
| **Second subunit** | | | | | | | | | | | | | |
| ASN_334 | 1 | | | | | | | | | | | | 1 |
| CYX_336 | | | | | 1 | | | | | | | | 1 |
| PHE_338 | | | | | 1 | | | | | | | | 1 |
| GLY_339 | | | | | 1 | | | | | | | | 1 |
| GLU_340 | | | | | 1 | | | | | | | | 1 |
| THR_345 | | 1 | | | | | | 1 | 1 | | | 1 | 4 |
| ARG_346 | 1 | 1 | 1 | 1 | 1 | | | 1 | 1 | 1 | | 1 | 9 |
| SER_349 | 1 | | | | | | | | | | | | 1 |
| TYR_351 | 1 | | | | | | | | | | | | 1 |

| Residue | 1 | 2 | 3 | 4 | 5 | 6 | 7 | 8 | 9 | 10 | 11 | Total |
|---|---|---|---|---|---|---|---|---|---|---|---|---|
| ASN_354 | | | 1 | 1 | 1 | | 1 | 1 | | | | 5 |
| ARG_355 | | | 1 | 1 | 1 | 1 | | | | | | 4 |
| LYS_356 | 1 | | 1 | | | | | | | | | 2 |
| ARG_357 | 1 | 1 | 1 | 1 | 1 | 1 | | 1 | | | 1 | 8 |
| SER_359 | 1 | | 1 | 1 | | | 1 | | 1 | | 1 | 6 |
| ASN_360 | 1 | | 1 | | 1 | 1 | 1 | | 1 | 1 | 1 | 8 |
| ASP_364 | | | | 1 | | | | | | | | 1 |
| SER_443 | | | 1 | | | | | | | | | 1 |
| LYS_444 | | | 1 | | | | | | 1 | | | 2 |
| VAL_445 | | | 1 | | 1 | | | | 1 | | | 3 |
| GLY_447 | | | 1 | | 1 | | | | 1 | | | 3 |
| ASN_448 | | 1 | | | | | | | | | | 1 |
| TYR_449 | | | | | 1 | | | | | | | 1 |
| ASN_450 | 1 | 1 | | | 1 | | | 1 | | | | 4 |
| LEU_452 | 1 | | | | | | | | | | | 1 |
| ARG_466 | | | | | 1 | | | | | | | 1 |
| LYS_558 | | | | 1 | | | | | | | | 1 |
| GLN_564 | | | | 1 | | | | 1 | | | | 2 |
| ARG_577 | | | 1 | | | | | 1 | | | | 2 |
| THR_581 | | | | 1 | | | | | | | 1 | 2 |

**Tab. S6 Interaction Fingerprint Analysis (MD-IFP) (3) between heparin and spike residues.** In Model02, one heparin chain interacts with residues of $S_A$ and $S_B$ subunits. In Models 03 and 05, each of the three heparin chains interacts with residues of $S_A$-$S_B$ or $S_B$-$S_C$ or $S_C$-$S_A$. Histograms show the interaction in the first 10 frames, last 10 frames and in all the frames in blue, orange and with lines, respectively. Blue and red boxes include the NTD-RBD and S1/S2 domains, respectively. To read the residue name and number, see the attached files IFP-model02, IFP-model03, IFP-model05. These files report the Amber numbering as follows. subunit $S_A$ residues 2-1124, subunit SB residues 1289-2411, subunit SC residue number 2574-3696 (fasta numbering for all the subunits; residues 13-1139).

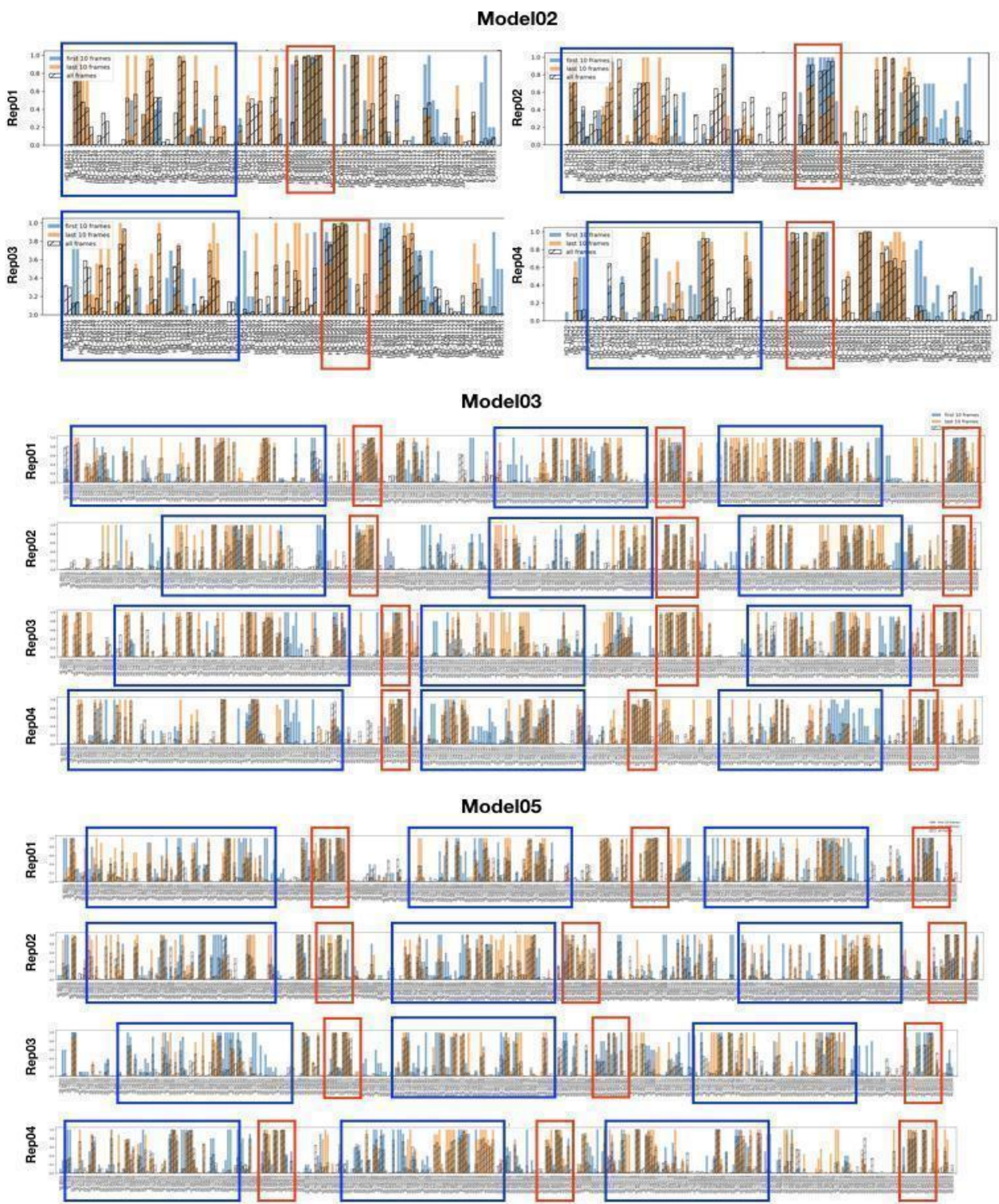

**Tab. S7** Solvent Accessible Surface Area (SASA) ($Å^2$) of the S1/S2 site (residues 682-685) calculated for the most representative cluster for each trajectory using NACCESS (4) with a probe radius of 1.4 Å.

| | subunit SA | Rep 01 | Rep 02 | Rep 03 | Rep 04 | AVG [$Å^2$] | ST.DEV. [$Å^2$] |
|---|---|---|---|---|---|---|---|
| **SASA S1/S2 Model 01** | SASA w/o glycans | 629.83 | 633.49 | 530.64 | 372.34 | 541.575 | 122.47 |
| | SASA w/ glycans | 566.64 | 633.34 | 530.64 | 328.5 | 514.78 | 131.27 |
| | Glycan shielded Area | 63.19 | 0.15 | 0 | 43.84 | 26.795 | 31.85 |
| | **subunit SB** | **Rep 01** | **Rep 02** | **Rep 03** | **Rep 04** | **AVG [$Å^2$]** | **ST.DEV. [$Å^2$]** |
| | SASA w/o glycans | 505.5 | 582.18 | 642.34 | 617.75 | 586.9425 | 59.65 |
| | SASA w/ glycans | 505.5 | 547.92 | 642.34 | 617.75 | 578.3775 | 62.93 |
| | Glycan shielded Area | 0 | 34.26 | 0 | 0 | 8.565 | 17.13 |
| | **subunit SC** | **Rep 01** | **Rep 02** | **Rep 03** | **Rep 04** | **AVG [$Å^2$]** | **ST.DEV. [$Å^2$]** |
| | SASA w/o glycans | 416.62 | 579.49 | 411.38 | 550.91 | 489.6 | 88.08 |
| | SASA w/ glycans | 390.87 | 579.49 | 391.61 | 535.57 | 474.385 | 97.67 |
| | Glycan shielded Area | 25.75 | 0 | 19.77 | 15.34 | 15.215 | 11.01 |
| | **subunit SA** | **Rep 01** | **Rep 02** | **Rep 03** | **Rep 04** | **AVG [$Å^2$]** | **ST.DEV. [$Å^2$]** |
| **SASA S1/S2 Model 02** | SASA w/o glycans- hepa | 629.66 | 642.73 | 453.62 | 600.37 | 581.595 | 87.14 |
| | SASA w/ glycans | 570.35 | 571.14 | 438.71 | 493.11 | 518.3275 | 64.47 |
| | SASA w/ glycans + hepa | 281.36 | 315.2 | 239.59 | 239.2 | 268.8375 | 36.70 |
| | Heparin shielded Area | 288.99 | 255.94 | 199.12 | 253.91 | 249.49 | 37.23 |
| | **subunit SB** | **Rep 01** | **Rep 02** | **Rep 03** | **Rep 04** | **AVG [$Å^2$]** | **ST.DEV. [$Å^2$]** |
| | SASA w/o glycans | 630.78 | 406.93 | 554.48 | 468.94 | 515.2825 | 97.92 |
| | SASA w/ glycans | 540.8 | 382.68 | 296.33 | 349.33 | 392.285 | 105.20 |
| | Glycan shielded Area | 89.98 | 24.25 | 258.15 | 119.61 | 122.9975 | 98.52 |
| | **subunit SC** | **Rep 01** | **Rep 02** | **Rep 03** | **Rep 04** | **AVG [$Å^2$]** | **ST.DEV. [$Å^2$]** |
| | SASA w/o glycans | 442.81 | 417 | 656.9 | 653.61 | 542.58 | 130.54 |
| | SASA w/ glycans | 350.95 | 417 | 598.6 | 653.61 | 505.04 | 144.13 |
| | Glycan shielded Area | 91.86 | 0 | 58.3 | 0 | 37.54 | 45.46 |
| | **subunit SA** | **Rep 01** | **Rep 02** | **Rep 03** | **Rep 04** | **AVG [$Å^2$]** | **ST.DEV. [$Å^2$]** |
| | SASA w/o glycans- hepa | 516.39 | 515.64 | 581.44 | 748.69 | 590.54 | 109.85 |
| | SASA w/ glycans | 487.64 | 426.44 | 563.91 | 746.23 | 556.055 | 138.65 |
| | SASA w/ glycans + hepa | 334.23 | 255.98 | 391.9 | 526.14 | 377.0625 | 113.93 |
| | Heparin shielded Area | 153.41 | 170.46 | 172.01 | 220.09 | 178.9925 | 28.66 |
| | **subunit SB** | **Rep 01** | **Rep 02** | **Rep 03** | **Rep 04** | **AVG [$Å^2$]** | **ST.DEV. [$Å^2$]** |
| **SASA S1/S2 Model 03** | SASA w/o glycans- hepa | 479.39 | 579.68 | 464.96 | 388.36 | 478.0975 | 78.62 |
| | SASA w/ glycans | 479.39 | 579.68 | 464.96 | 388.36 | 478.0975 | 78.62 |
| | SASA w/ glycans + hepa | 293.57 | 380.35 | 244.35 | 189.05 | 276.83 | 81.15 |
| | Heparin shielded Area | 185.82 | 199.33 | 220.61 | 199.31 | 201.2675 | 14.38 |
| | **subunit SC** | **Rep 01** | **Rep 02** | **Rep 03** | **Rep 04** | **AVG [$Å^2$]** | **ST.DEV. [$Å^2$]** |
| | SASA w/o glycans- hepa | 418.24 | 567.65 | 609.42 | 588.14 | 545.8625 | 86.77 |
| | SASA w/ glycans | 418.24 | 567.65 | 574.51 | 557.84 | 529.56 | 74.53 |
| | SASA w/ glycans + hepa | 257.57 | 388.05 | 387.46 | 340.88 | 343.49 | 61.39 |
| | Heparin shielded Area | 160.67 | 179.6 | 187.05 | 216.96 | 186.07 | 23.40 |
| | **subunit SA** | **Rep 01** | **Rep 02** | **Rep 03** | **Rep 04** | **AVG [$Å^2$]** | **ST.DEV. [$Å^2$]** |
| | SASA w/o glycans | 513.33 | 530.3 | 504.07 | 577.56 | 531.315 | 32.683 |
| | SASA w/ glycans | 513.33 | 530.3 | 384.19 | 577.56 | 501.345 | 82.70 |
| | Glycan shielded Area | 0 | 0 | 119.88 | 0 | 29.97 | 59.94 |
| **SASA S1/S2 Model 04** | **subunit SB** | **Rep 01** | **Rep 02** | **Rep 03** | **Rep 04** | **AVG [$Å^2$]** | **ST.DEV. [$Å^2$]** |
| | SASA w/o glycans | 537.96 | 490.5 | 387.01 | 426.75 | 460.555 | 66.93 |
| | SASA w/ glycans | 468.25 | 490.5 | 387.01 | 323.6 | 417.34 | 76.70 |
| | Glycan shielded Area | 69.71 | 0 | 0 | 103.15 | 43.215 | 51.73 |
| | **subunit SC** | **Rep 01** | **Rep 02** | **Rep 03** | **Rep 04** | **AVG [$Å^2$]** | **ST.DEV. [$Å^2$]** |
| | SASA w/o glycans | 287.33 | 450.2 | 373.27 | 571.4 | 420.55 | 120.57 |
| | SASA w/ glycans | 268.55 | 450.2 | 373.27 | 571.4 | 415.855 | 127.65 |
| | Glycan shielded Area | 18.78 | 0 | 0 | 0 | 4.695 | 9.39 |

| | | Rep 01 | Rep 02 | Rep 03 | Rep 04 | AVG [Å²] | ST.DEV. [Å²] |
|---|---|---|---|---|---|---|---|
| **SASA S1/S2 Model 05** | **subunit SA** | **Rep 01** | **Rep 02** | **Rep 03** | **Rep 04** | **AVG [Å²]** | **ST.DEV. [Å²]** |
| | SASA w/o glycans- hepa | 586.63 | 608.31 | 620.32 | 473.41 | 572.1675 | 67.30 |
| | SASA w/ glycans | 586.63 | 607.13 | 620.32 | 473.41 | 571.8725 | 67.09 |
| | SASA w/ glycans + hepa | 240.61 | 273.7 | 410.65 | 281.39 | 301.5875 | 74.83 |
| | Heparin shielded Area | 346.02 | 333.43 | 209.67 | 192.02 | 270.285 | 80.67 |
| | **subunit SB** | **Rep 01** | **Rep 02** | **Rep 03** | **Rep 04** | **AVG [Å²]** | **ST.DEV. [Å²]** |
| | SASA w/o glycans- hepa | 554.89 | 440.05 | 562.46 | 378.39 | 483.9475 | 89.94 |
| | SASA w/ glycans | 499.14 | 440.05 | 554.17 | 378.39 | 467.9375 | 75.73 |
| | SASA w/ glycans + hepa | 203.26 | 381.9 | 438.86 | 224.4 | 312.105 | 116.16 |
| | Heparin shielded Area | 295.88 | 58.15 | 115.31 | 153.99 | 155.8325 | 101.32 |
| | **subunit SC** | **Rep 01** | **Rep 02** | **Rep 03** | **Rep 04** | **AVG [Å²]** | **ST.DEV. [Å²]** |
| | SASA w/o glycans- hepa | 481.46 | 424.66 | 559.17 | 537.64 | 500.7325 | 60.37 |
| | SASA w/ glycans | 456.98 | 424.66 | 509.52 | 537.64 | 482.2 | 50.88 |
| | SASA w/ glycans + hepa | 183.05 | 203.28 | 293.16 | 320.88 | 250.0925 | 67.21 |
| | Heparin shielded Area | 273.93 | 221.38 | 216.36 | 216.76 | 232.1075 | 27.97 |

**Tab. S8 Furin cleavage assay.** Single measurements obtained for the various experimental conditions used to investigate the ability of heparin to inhibit furin cleavage by binding at the spike S1/S2 site.

|  | Rep 1 | Rep 2 | Rep 3 | Rep 4 | Rep 5 | Rep 6 | Rep 7 | Rep 8 | Rep 9 | Rep 10 | AVG | ST. DEV. |
|---|---|---|---|---|---|---|---|---|---|---|---|---|
| **CTRL** | 0.191 | 0.222 | 0.214 | 0.228 | 0.219 | 0.239 | 0.431 | 0.405 | 0.264 | - | **0.268** | **0.087** |
| **Heparin (10 μM)** | 0.18 | 0.218 | 0.187 | 0.196 | - | - | - | - | - | - | **0.196** | **0.017** |
| **Furin (25ng/well)** | 0.014 | 0.052 | 0.027 | 0.027 | 0.054 | 0.075 | 0.075 | 0.123 | 0.046 | 0.045 | **0.054** | **0.031** |
| **Furin (25ng/well) + Heparin (10 μM)** | 0.338 | 0.277 | 0.325 | 0.337 | - | - | - | - | - | - | **0.319** | **0.029** |
| **Heparin (10 μM) → PBS → Furin (25ng/well)** | 0.331 | 0.31 | 0.322 | 0.229 | - | - | - | - | - | - | **0.298** | **0.047** |
| **Heparin (10 μM) → Salt (2M NaCl) → Furin (25ng/well)** | 0.111 | 0.144 | 0.098 | 0.054 | - | - | - | - | - | - | **0.102** | **0.037** |

**Tab. S9 Furin cleavage assay.** Single measurements obtained to compare the ability of unfractionated heparin (UFH), low molecular weight heparin (LMWH) and unsulfated K5 polysaccharide (K5) to inhibit furin cleavage of the spike S1/S2 site. In each measurement 25ng/well of furin was used.

|  | Concentration [μM] | Rep 1 | Rep 2 | Rep 3 | Rep 4 | Rep 5 | Rep 6 | AVG | ST. DEV. |
|---|---|---|---|---|---|---|---|---|---|
| **UFH** | **0.001** | 0.038 | 0.008 | 0.01 | - | - | - | **0.019** | **0.017** |
|  | **0.01** | 0.137 | 0.135 | 0.207 | 0.17 | 0.074 | 0.046 | **0.13** | **0.06** |
|  | **0.1** | 0.228 | 0.245 | 0.158 | 0.098 | - | - | **0.182** | **0.068** |
|  | **1** | 0.203 | 0.226 | 0.238 | 0.242 | - | - | **0.227** | **0.018** |
|  | **10** | 0.338 | 0.277 | 0.325 | 0.337 | - | - | **0.319** | **0.029** |
|  | **100** | 0.32 | 0.328 | 0.317 | - | - | - | **0.317** | **0.006** |
| **LMWH** | **0.01** | 0.055 | 0.109 | 0.08 | - | - | - | **0.0813** | **0.027** |
|  | **0.1** | 0.099 | 0.076 | 0.091 | - | - | - | **0.087** | **0.012** |
|  | **1** | 0.164 | 0.176 | 0.117 | 0.152 | - | - | **0.152** | **0.025** |
|  | **10** | 0.152 | 0.145 | 0.193 | 0.197 | - | - | **0.171** | **0.027** |
|  | **100** | 0.255 | 0.301 | 0.279 | - | - | - | **0.278** | **0.023** |
|  | **1000** | 0.293 | 0.287 | 0.288 | - | - | - | **0.289** | **0.003** |
| **K5** | **10** | 0.037 | 0.021 | 0.025 | 0.015 | - | - | **0.024** | **0.009** |

**Tab. S10** Solvent Accessible Surface Area (SASA) (Å$^2$) of residues involved in binding with the ACE2 receptor extracted from the crystallographic data of Lan et al. (5) and calculated for the most representative cluster using NACCESS (4) with a probe radius of 1.4 Å.

| | | Rep 01 | Rep 02 | Rep 03 | Rep 04 | AVG [Å$^2$] | ST.DEV. [Å$^2$] |
|---|---|---|---|---|---|---|---|
| **SASA RBD Model 01** | **subunit SA** | | | | | | |
| | SASA w/o glycans | 1535.01 | 1504.43 | 1210.01 | 1260.4 | 1377.4625 | 166.02 |
| | SASA w/ glycans | 1311.63 | 1122.36 | 939.32 | 1126.82 | 1125.0325 | 152.01 |
| | Glycan shielded Area | 223.38 | 382.07 | 270.69 | 133.58 | 252.43 | 103.46 |
| | **subunit SB** | **Rep 01** | **Rep 02** | **Rep 03** | **Rep 04** | **AVG [Å$^2$]** | **ST.DEV. [Å$^2$]** |
| | SASA w/o glycans | 1332.19 | 1435.41 | 1302.66 | 1193.93 | 1316.0475 | 99.33 |
| | SASA w/ glycans | 1040.84 | 1039.22 | 781.12 | 740.21 | 900.3475 | 162.16 |
| | Glycan shielded Area | 291.35 | 396.19 | 521.54 | 453.72 | 415.7 | 97.46 |
| | **subunit SC** | **Rep 01** | **Rep 02** | **Rep 03** | **Rep 04** | **AVG [Å$^2$]** | **ST.DEV. [Å$^2$]** |
| | SASA w/o glycans | 1144.36 | 1316.63 | 1568.52 | 1352.71 | 1345.555 | 174.24 |
| | SASA w/ glycans | 793.47 | 1216.78 | 1255.32 | 1267.56 | 1133.2825 | 227.57 |
| | Glycan shielded Area | 350.89 | 99.85 | 313.2 | 85.15 | 212.2725 | 139.28 |
| **SASA RBD Model 02** | **subunit SA** | **Rep 01** | **Rep 02** | **Rep 03** | **Rep 04** | **AVG [Å$^2$]** | **ST.DEV. [Å$^2$]** |
| | SASA w/o glycans | 1228.63 | 1201.1 | 1505 | 1519.19 | 1363.48 | 172.07 |
| | SASA w/ glycans | 779.14 | 1115 | 1222.72 | 1369.44 | 1121.575 | 250.98 |
| | Glycan shielded Area | 449.49 | 86.1 | 282.28 | 149.75 | 241.905 | 160.72 |
| | **subunit SB** | **Rep 01** | **Rep 02** | **Rep 03** | **Rep 04** | **AVG [Å$^2$]** | **ST.DEV. [Å$^2$]** |
| | SASA w/o glycans- hepa | 1417.49 | 1206.9 | 1274.83 | 1329.75 | 1307.2425 | 89.03 |
| | SASA w/ glycans | 1042.8 | 985.55 | 1200.53 | 1047.38 | 1069.065 | 92.05 |
| | SASA w/ glycans + hepa | 1042.8 | 985.55 | 1200.53 | 1047.38 | 1069.065 | 92.05 |
| | Heparin shielded Area | 374.69 | 221.35 | 74.3 | 282.37 | 238.1775 | 126.13 |
| | **subunit SC** | **Rep 01** | **Rep 02** | **Rep 03** | **Rep 04** | **AVG [Å$^2$]** | **ST.DEV. [Å$^2$]** |
| | SASA w/o glycans | 1369.09 | 1201.23 | 1330.53 | 1335.14 | 1308.9975 | 73.87 |
| | SASA w/ glycans | 900.1 | 791.89 | 963.62 | 1296.57 | 988.045 | 217.56 |
| | Glycan shielded Area | 468.99 | 409.34 | 366.91 | 38.57 | 320.9525 | 192.86 |
| **SASA RBD Model 03** | **subunit SA** | **Rep 01** | **Rep 02** | **Rep 03** | **Rep 04** | **AVG [Å$^2$]** | **ST.DEV. [Å$^2$]** |
| | SASA w/o glycans- hepa | 1416.74 | 1050.74 | 921.89 | 1412.17 | 1200.385 | 252.73 |
| | SASA w/ glycans | 1170.23 | 671.73 | 774.8 | 1188.74 | 951.375 | 266.85 |
| | SASA w/ glycans + hepa | 1168.11 | 659.13 | 774.8 | 1136.32 | 934.59 | 256.02 |
| | Heparin shielded Area | 2.12 | 12.6 | 0 | 52.42 | 16.785 | 24.39 |
| | Glycan shielded Area | 248.63 | 391.61 | 147.09 | 275.85 | 249.01 | 100.53 |
| | **subunit SB** | **Rep 01** | **Rep 02** | **Rep 03** | **Rep 04** | **AVG [Å$^2$]** | **ST.DEV. [Å$^2$]** |
| | SASA w/o glycans- hepa | 1009.43 | 993.37 | 1108.68 | 1135.53 | 1061.7525 | 70.85 |
| | SASA w/ glycans | 850.01 | 739.51 | 687.97 | 886.77 | 791.065 | 92.95 |
| | SASA w/ glycans + hepa | 831.2 | 739.51 | 687.97 | 868.53 | 781.8025 | 82.77 |
| | Heparin shielded Area | 18.81 | 0 | 0 | 18.24 | 9.2625 | 10.70 |
| | Glycan shielded Area | 178.23 | 253.86 | 420.71 | 267 | 279.95 | 101.67 |
| | **subunit SC** | **Rep 01** | **Rep 02** | **Rep 03** | **Rep 04** | **AVG [Å$^2$]** | **ST.DEV. [Å$^2$]** |
| | SASA w/o glycans- hepa | 1266.88 | 1188.41 | 1374.5 | 1259.85 | 1272.41 | 76.74 |
| | SASA w/ glycans | 1218.42 | 1000.93 | 1097.99 | 1015.93 | 1083.3175 | 99.66 |
| | SASA w/ glycans + hepa | 1218.42 | 1000.93 | 1097.99 | 1015.93 | 1083.3175 | 99.66 |
| | Heparin shielded Area | 0 | 0 | 0 | 0 | 0 | 0 |
| | Glycan shielded Area | 48.46 | 187.48 | 276.51 | 243.92 | 189.0925 | 100.71 |
| **SASA RBD Model 04** | **subunit SA** | **Rep 01** | **Rep 02** | **Rep 03** | **Rep 04** | **AVG [Å$^2$]** | **ST.DEV. [Å$^2$]** |
| | SASA w/o glycans | 934.02 | 1023.2 | 1331.32 | 925.07 | 1053.40 | 190.50 |
| | SASA w/ glycans | 925.07 | 1023.2 | 1331.32 | 925.07 | 1051.165 | 192.41 |
| | Glycan shielded Area | 8.95 | 0 | 0 | 0 | 2.23 | 4.475 |
| | **subunit SB** | **Rep 01** | **Rep 02** | **Rep 03** | **Rep 04** | **AVG [Å$^2$]** | **ST.DEV. [Å$^2$]** |
| | SASA w/o glycans | 1294.16 | 1302.6 | 1352.26 | 1294.16 | 1310.795 | 27.92 |
| | SASA w/ glycans | 1115.78 | 1302.6 | 1029.59 | 1115.78 | 1140.94 | 115.18 |
| | Glycan shielded Area | 178.38 | 0 | 322.67 | 178.38 | 169.86 | 132.10 |
| | **subunit SC** | **Rep 01** | **Rep 02** | **Rep 03** | **Rep 04** | **AVG [Å$^2$]** | **ST.DEV. [Å$^2$]** |
| | SASA w/o glycans | 1562.47 | 1302.6 | 1546.65 | 1562.47 | 1493.5475 | 127.52 |

|  | | Rep 01 | Rep 02 | Rep 03 | Rep 04 | AVG [Å²] | ST.DEV. [Å²] |
|---|---|---|---|---|---|---|---|
|  | SASA w/ glycans | 1562.41 | 1302.6 | 1545.93 | 1562.41 | 1493.3375 | 127.40 |
|  | Glycan shielded Area | 0.06 | 0 | 0.72 | 0.06 | 0.21 | 0.34 |
| **SASA RBD Model 05** | **subunit SA** | **Rep 01** | **Rep 02** | **Rep 03** | **Rep 04** | **AVG [Å²]** | **ST.DEV. [Å²]** |
|  | SASA w/o glycans- hepa | 811.23 | 717.19 | 851.27 | 824.47 | 801.04 | 58.33 |
|  | SASA w/ glycans | 726.13 | 516.61 | 802.55 | 679.78 | 681.2675 | 120.88 |
|  | SASA w/ glycans + hepa | 726.13 | 504.51 | 802.55 | 620.42 | 663.4025 | 129.60 |
|  | Heparin shielded Area | 0 | 12.1 | 0 | 59.36 | 17.865 | 28.25 |
|  | Glycan shielded Area | 85.1 | 212.68 | 48.72 | 204.05 | 137.6375 | 83.08 |
|  | **subunit SB** | **Rep 01** | **Rep 02** | **Rep 03** | **Rep 04** | **AVG [Å²]** | **ST.DEV. [Å²]** |
|  | SASA w/o glycans- hepa | 1112.45 | 993.19 | 1167.5 | 1455.88 | 1139.975 | 196.39 |
|  | SASA w/ glycans | 778.58 | 786.46 | 993.76 | 1233.6 | 890.11 | 214.83 |
|  | SASA w/ glycans + hepa | 778.58 | 786.46 | 981.52 | 1233.6 | 883.99 | 214.05 |
|  | Heparin shielded Area | 0 | 0 | 12.24 | 0 | 0 | 6.12 |
|  | Glycan shielded Area | 333.87 | 206.73 | 185.98 | 222.28 | 214.505 | 66.13 |
|  | **subunit SC** | **Rep 01** | **Rep 02** | **Rep 03** | **Rep 04** | **AVG [Å²]** | **ST.DEV. [Å²]** |
|  | SASA w/o glycans- hepa | 1512.39 | 1384.55 | 1370.12 | 1356.64 | 1405.925 | 71.89 |
|  | SASA w/ glycans | 1512.39 | 1384.55 | 1370.12 | 1356.64 | 1405.925 | 71.89 |
|  | SASA w/ glycans + hepa | 1512.39 | 1351.2 | 1370.12 | 1239.3 | 1368.2525 | 112.10 |
|  | Heparin shielded Area | 0 | 33.35 | 0 | 117.34 | 37.6725 | 55.39 |
|  | Glycan shielded Area | 0 | 33.35 | 0 | 117.34 | 37.6725 | 55.39 |

## 4. Supplementary Movies

**Movie S1. Spike glycoprotein in a closed conformation with 1 heparin chain bound.**
First the structure is shown, focusing on the heparin binding site, then the MD trajectory for replica 1 (corresponding to the figures in the main manuscript).

**Movie S2. Spike glycoprotein in a closed conformation with 3 heparin chains bound.**
The structure and how the three heparin chains bind is shown followed by the MD trajectory for replica 1 (corresponding to the figures in the main manuscript).

**Movie S3. Spike glycoprotein in an open conformation with 3 heparin chains bound.**
The structure and how the three heparin chains bind is shown, focusing on the binding to the open RBD, followed by the MD trajectory for replica 1 (corresponding to the figures in the main manuscript).

## 5. Supplementary References